%% file: main.tex
\documentclass{llncs}
\pdfoutput=1
\usepackage[utf8]{inputenc}
\usepackage{style}

\title{Fundamental Machine Learning Routines as Quantum Algorithms on a Superconducting Quantum Computer
}

\author{Sristy Sangskriti\inst{1}\textsuperscript{*} 
\and
Protik Nag\inst{1}
\and
Summit Haque\inst{1}
}
\institute{Shahjalal University of Science and
Technology\\ Kumargaon, Sylhet 3114, Bangladesh\\
\email{sristysangskriti007@gmail.com}\\
 }
\begin{document}

\makeatletter
\renewcommand*\l@author[2]{}
\renewcommand*\l@title[2]{}
\makeatletter

 \maketitle

\pagestyle{plain}

\pagenumbering{roman}

\begin{abstract}
\addcontentsline{toc}{section}{Abstract}
 The Harrow-Hassidim-Lloyd algorithm is intended for solving the system of linear equations on quantum devices. The exponential advantage of the algorithm comes with four caveats. We present a numerical study of the performance of the algorithm when these caveats are not perfectly matched. We observe that, between diagonal and non-diagonal matrices, the algorithm performs with higher success probability for the diagonal matrices. At the same time, it fails to perform well on lower or higher density sparse Hermitian matrices. Again, Quantum Support Vector Machine algorithm is a promising algorithm for classification problem. We have found out that it works better with binary classification problem than multi-label classification problem. And there are many opportunities left for improving the performance. 

\end{abstract}
\clearpage

\addcontentsline{toc}{section}{Table of Contents}
\tableofcontents

\newpage

\pagenumbering{arabic}

\section{Introduction}
\input{tex/intro}

\section{Statement of the Problems and Summary of Results}
\input{tex/problem_statement}

\section{Preliminary Concepts}
\input{tex/review}

\section{Literature Review}

\input{tex/foundation}

\section{The IBM Quantum Computing Platform}
\input{tex/ibm}

\section{Design of the experiments}
\input{tex/experiment}

\section{Results and Discussion}

\input{tex/result}

\section{Future Work}

\input{tex/future}

\section{Conclusion}
\input{tex/conclusion.tex}

\section{Acknowledgements}
\input{tex/acknowledgement} 

\section{Dedication}
\input{tex/dedication} 
%
%
%
\bibliographystyle{splncs04}
\bibliography{ref}

\end{document}

%% file: tex/intro.tex
\label{intro}
 As humans become increasingly dependent on machines, we are observers to a modern revolution that is assuming control over the world, and that will be the inevitable future of Machine Learning. Machine learning algorithms are developed in such a way that they can process lots of data together and train a model to make decisions. But in order to speed up this process we need a new technology called quantum computers which will make revolutionary change in this field.\\
Quantum computing, an applied science of quantum mechanical phenomena, was introduced with the notion of simulating things that a traditional computer can not do. Quantum computers are accelerated machines that can compute exponentially faster than classical computers. New algorithms for solving difficult problems using quantum computers have been being developed for years. One of these problems are systems of linear equations.\\
Quantum computer can be a life saver in the field of machine learning and the vast of AI deserves a big change. Quantum machine learning will help promote the computational power to handle problems that are too complex too solve. Some discussable fields of quantum machine learning algorithms are finding eigenvalues and eigenvectors of large matrices, analysis of classical machine learning algorithms on quantum computers etc.\\
Besides, traditional computers increases their processing power by expanding the quantity of transistors which are the essential part of the basic structure of classical computers. A transistor can just hold one state at a time i.e. switch on or off, an ever increasing number of transistors  is necessary to process on every single possible states. So the best way to amplify computing power is to increase the number of transistors and to minify them. The mostly developed chips today consist of a huge number of transistors on a single chip. But there is also a limitation to how little they can get. Till now there is no quantum effect of their sizes but someday it will impossible to make transistors any smaller \cite{samee2018quantum}.
System of linear equations is of great importance in basically every field of science and engineering and also in finance, economics. These systems seem to appear in the area of solving differential or partial equations and also in machine learning problems such as regression. The HHL algorithm can be used as a subroutine in a bigger quantum algorithm. It is a fundamental algorithm which can further be used to support many quantum machine algorithms\cite{biamonte2017quantum,dunjko2017machine,aaronson2015read, rebentrost2014quantum,rebentrost2018quantum, schuld2016prediction, wiebe2012quantum}.  But due to the ever increasing amount of data, it becomes too difficult to acquire a solution by processing this huge amount of data. This situation requires a great deal of computational power and precious time. \\
Although quantum computers can accelerate the processing for some problems than classical ones, there are many parts which still contain some uncertainty or restrictions \cite{shao2018reconsider}. We can mention the quantum phase estimation algorithm \cite{kitaev1995quantum} in this context. The eigenvalue estimation of hermitian matrix is one major step of the HHL algorithm. Hermitian matrix is discussed in the latter chapter. As the phase estimation system is embedded in this algorithm there are also some uncertainty of getting the exact output from this algorithm. \\
But quantum computers may not supplant classical computers. Quantum computers are supposed to be able to work on specific kinds of problems like large optimizations. They will also be expensive to run. But quantum computers have exponential potential over classical computers. When classical computers use backtracking in order to find a path through a tangle, quantum computer parellelly search through every possible path to find the destination. By adding more qubits to a quantum computer  the power of the computer can be increased at an exponential rate.\\
Machine Learning algorithms such as neural networks sometimes require too much time to train due to lack of computational power. It has become a necessity to accelerate the training process and making complex algorithms feasible. To solve such problems related to computational complexity, quantum computing is introduced. Using Support Vector Machine algorithms on quantum computers for helpful classification of data and dealing with complex problems may be a small step. \\
Quantum computers can be used to do prime factorization in no time and this unleash demolition on cyber security. Quantum computing can also be applied to understand nano particles, designing chemicals and drugs, discovery about the universe, in renewable energy and so on. It may revolutionize every aspect of our society.
Machine learning  is a common platform to almost every field of science because it helps to make decisions and gives an overview of the scenery. But the base of machine learning lies in mathematics. The field of mathematics has reached beyond the scope classical mechanics and mathematics gives new ideas to improve machine learning algorithms. But to implement these algorithms and ideas scientists need greater computational power which classical computers can not meet always  or even if they can process the ideas taking too much time. So an era of quantum computers is coming toward us.

%% file: tex/problem_statement.tex

\label{statement}
This whole thesis work can be split into two major parts. First we summarize the statements and then give a summary of the results.

\subsection{A Numerical Study of the Harrow-Hassidim-Lloyd Algorithm for Solving Systems of Linear Equations} 
The Harrow-Hassidim-Lloyd algorithm is used to solve systems of linear equations. The details about this algorithm is described in the latter chapter \ref{review}. In our contribution, we have decided to benchmark this algorithm. We have tried to find out the flaws and have provided some corner cases of this algorithm. We have measured the output it returns at different cases. Our goal of this contribution is to discover how the  HHL algorithm performs on various matrices. The algorithm has been examined with different sizes of matrices, with differently dense matrices and with both diagonal and non-diagonal matrices. We have learnt that the HHL algorithm performs well with small sized, low density, diagonal matrices. Otherwise, the performance starts to decline. Chapter \ref{experiment} contains more details about the experiment we have conducted to come to this conclusion. 

\subsection{ Comparative Study of Different Quantum Machine Learning Algorithm for Solving a Binary Classification Problem }

Support Vector Machine (SVM) is a traditional classical machine learning approach, where vectors are constructed through the problem space such that they divide up the problem data into separate classifications. We expect quantum computers to be able to determine those vectors faster than classical hardware. The reason behind this expectation is explained in the chapter \ref{review}. Though, quantum hardwares are not at a stage where the performance outpaces that of classical hardwares currently. Our contribution shows results of an experiment of implementing the quantum support vector machine algorithm for a binary classification problem and for a multiple class problem and analyze those results with the classical ones.

\subsection{ Summary of Results }
In a nutshell, we can divide our research contribution into two parts. One is to benchmark the HHL algorithm and another one is to run a quantum machine learning algorithm on a standard dataset and correlate the results with a classical one. 

At first, while working on the HHL algorithm, we have observed some key points. We have found out that this algorithm performs better with smaller-sized matrices and the performance declines dramatically when the number of equations increases. Also, it functions better with lower density matrices. As the density increases, both performance and efficiency reduces. All the graphs are provided in chapter \ref{result}. Again, to solve higher dimensional system of linear equations, it requires larger circuit, which likewise reduces the productivity and becomes more susceptible to inaccuracy. Surprisingly, this algorithm works better on diagonal matrices than the non-diagonal ones. 

In the later part, we have experimented the quantum subpart of the Support Vector Machine algorithm. We have used two datasets named the Breast Cancer Dataset\cite{wolberg1992breast} and the  Iris Dataset\cite{fisher1936uci}. Description of these datasets are presented in the chapter \ref{experiment}. We have applied QSVM on these datasets and measured the accuracy. Also, we have run the classical SVM and gathered the results. We have studied both of the procedures and have found out that quantum devices do not perform well regarding Support Vector Machine algorithm yet. They are still far away from the state-of-the-art classical performance. But the performance can be stepped up by conducting the same experiments several times and taking the finest output. How the performance increases as the number of shots increases has been explained in chapter \ref{result}. Though this kind of approach can be inefficient in terms of memory and time.

%% file: tex/review.tex
\label{review}
Before we start to discuss about the quantum system and its components, we need some basic knowledge about some mathematical terms and theorems. These are complex numbers, qubits, superposition and so on. Then we discuss about the research of other people integrating machine learning on quantum computers.

\subsection{Complex Vector Space}
Complex number is the base of quantum computing. Complex number is the field of mathematics where lessons of quantum computing starts from. Although complex numbers were practised because of mathematical curiosity, scientists gradually understood that complex numbers are omnipresent and very much important. 

A \textit{complex vector space} is nonempty set of vectors $\mathbb{V}$ which satisfies addition, negation and scalar multiplication and a distinct element zero vector $0\in \mathbb{V}$ in the set \cite{yanofsky2008quantum}. 

If $\mathbb{B} = \{V_0,V_1,.....,V\textsubscript{n-1}\}$ is a set of vectors in complex vector space $\mathbb{V}$ so that every $V\in \mathbb{V}$ can be written as a linear combination of vectors from  $\mathbb{B}$ and $\mathbb{B}$ is linearly independent i.e. $0= c_0\cdot V_0+ c_1 \cdot V_1+.....+c\textsubscript{n-1}\cdot V\textsubscript{n-1}$ where $c_0=c_1=.....=c\textsubscript{n-1}=0$ , then $\mathbb{B} \in \mathbb{V}$ is called a \textit{basis} of $\mathbb{V}$. A \textit{canonical basis} or \textit{standard basis} is comprised of a number of functions $f_j(j=0,1,2,3,...)$ where $f_j =$  
$\begin{cases} 
      1, & if j = n, \\
      
      0, & otherwise
\end{cases}$. 

A change of basis matrix or \textit{transition matrix} is another matrix. It refers to the change from some basis $\mathbb{B}$ to some basis $\mathbb{E}$ for any vector $V$ such that $V_E = M\textsubscript{E$\leftarrow$ B} \cdot V_B $, where $M\textsubscript{E$\leftarrow$ B}$ is the transition matrix.

Now,  a \textit{Hadamard matrix} $H$ is the transition matrix from the canonical basis  $\Bigg{\{}\left[\begin{matrix}1\\0\end{matrix}\right], \left[\begin{matrix}0\\1\end{matrix}\right]\Bigg{\}}$  to this  other basis $\Bigg{\{}\left[\begin{matrix}\frac{1}{\sqrt{2}}\\\frac{1}{\sqrt{2}}\end{matrix}\right], \left[\begin{matrix}\frac{1}{\sqrt{2}}\\-\frac{1}{\sqrt{2}}\end{matrix}\right]\Bigg{\}}$ in $\mathbb{R}^2$. 

\begin{figure}[H]
     \centering
     \includegraphics[scale=0.5]{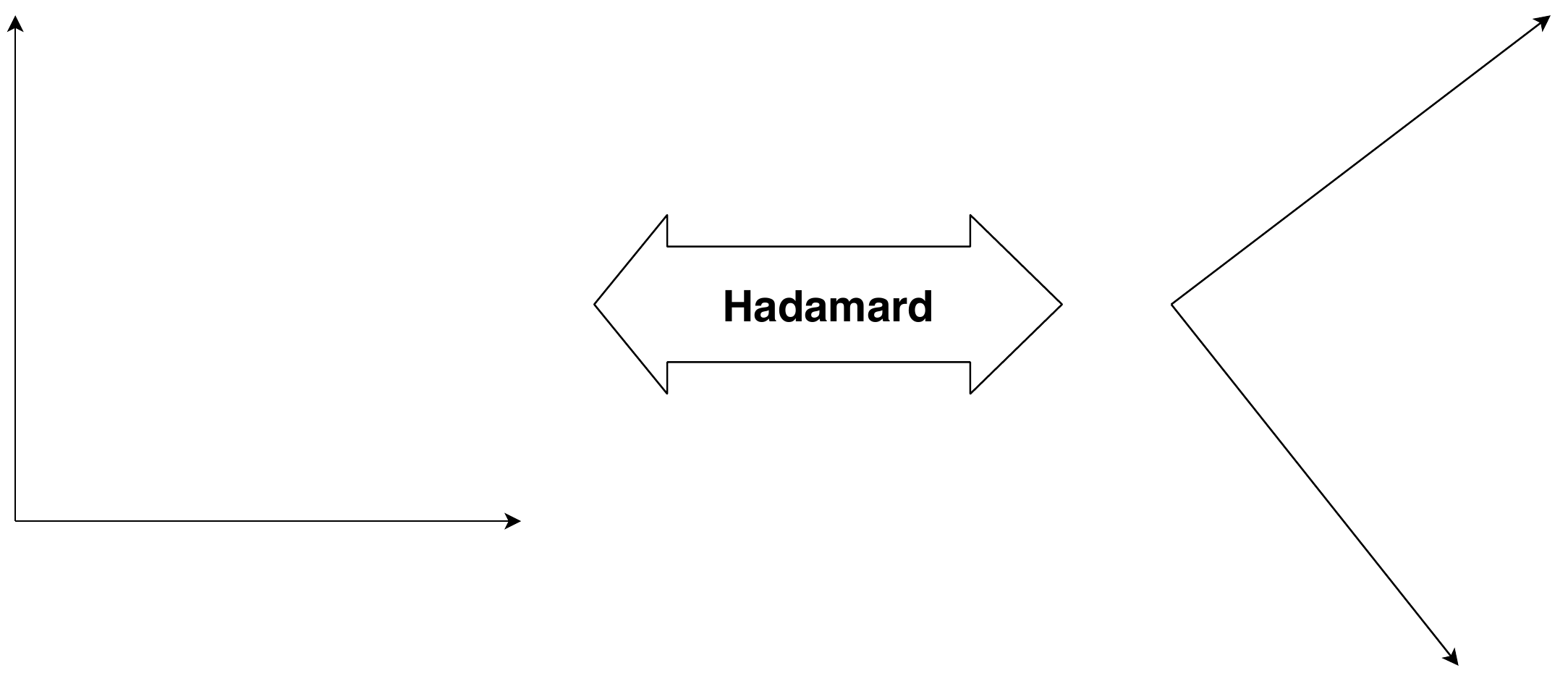}
     \caption{The Hadamard matrix shown as a transition between two bases\cite{yanofsky2008quantum}.}
     \label{fig:hadamard}
\end{figure}

In figure \ref{fig:hadamard} the transition between two hadamard matrices is depicted. 

A Hadamard matrix multiplied by itself gives the identity matrix. The transition back to the canonical basis is  a Hadamard matrix too. So Hadamard matrix has the property of  "reversibility". Hadamard matrices play a major role in quantum computing.It is used in data encryption, signal processing, data compression algorithms. It is a crucial part of Shor's algorithm\cite{shor1994algorithms}.

\subsection{Qubit and Superposition}

As depicted in the book of ``Quantum Computing For Computer Scientists" \cite{yanofsky2008quantum}, in classical computing system we measure information using \textit{bit}.  But in two dimensional quantum system information is measured by quantum bit or \textit{qubit}. In classical binary system there are two bits $0$ and $1$. These two bits can either indicate presence or absence of electricity or denote true or false or maybe refer to switch on or off. So a bit is a system where there are only two possible states. We can represent these states by 2-by-1 matrices such as state 
$\ket{0} =$ 
$\begin{blockarray}{cccc} 
&&& \\
            \begin{block}{ccc(c)}
                 0 & & & 1 \\
                 1 & & & 0 \\
\end{block}
\end{blockarray}$  
 \hspace{5 pt}   and 
 $\ket{1} =$ 
 $\begin{blockarray}{cccc}
&&& \\
\begin{block}{ccc(c)}
  0 & & & 0 \\
  1 & & & 1 \\
\end{block}
\end{blockarray}$   
\hspace{2 pt} .

Now in quantum world, a qubit can be in one state and in other state simultaneously i.e. a qubit can represent true and false or state $\ket{0}$ and  $\ket{1}$ at the same time.  We can represent such a qubit by a 2-by-1 matrix with complex numbers  
$\ket{1} =$ $\begin{blockarray}{cccc}
& & & \\
\begin{block}{ccc(c)}
  0 & & & c_{0} \\
  1 & & & c_{1} \\
\end{block}
\end{blockarray}$ \hspace{5 pt} 
where $\lvert c_0\rvert^2 + \lvert c_1\rvert^2 = 1$. A qubit becomes a bit after being measured. One of these $\lvert c_x\rvert^2$ is to be interpreted as the probability that after measuring the qubit it will be found in state $\ket{x}$ where $x=0,1$.

Thus one qubit represents two different states, two qubits can take on any of the four probable states 01, 11, 10, or 00. So $n$ qubits can represent $2^n$ different states. This concept of superposition is the key for quantum computers to be much more powerful than classical computers. The key idea is to represent much more information using less computing power. 

Let us suppose that a particle can be only be detected at one of the equally spaced points ${x_0,x_1,....,x\textsubscript{n-1}}$ of a line where $x_1=x_0+\delta x, x_2=x_1+\delta x,...$ with some fixed increment $\delta x$. The particle being at point $x_i$ is denoted as $\ket{x_i}$ (Dirac ket notation). Each of these basic states is associated with an $n$ dimensional complex column vector generally denoted as $[c_0,c_1,.....,c\textsubscript{n-1}]^T$. Now, an arbitrary state denoted as $\ket{\psi}$ is a linear combination of $\ket{x_0}, \ket{x_1},.....,\ket{x\textsubscript{n-1}}$ 
\begin{equation}
    \ket{\psi}= c_0\ket{x_0}+c_1\ket{x_1}+.....+c\textsubscript{n-1}\ket{x\textsubscript{n-1}}   
\end{equation}
where $c_0,c_1,....c\textsubscript{n-1}$ are complex amplitudes. This $\ket{\psi}$ can be described as the overlap of $n$ waves each contributing with amplitude $c_i$. We call this $\ket{\psi}$ a \textbf{superposition} of the basic states, which represents the particle as being in all locations ${x_0,x_1,....x\textsubscript{n-1}}$ at the same time.The probability of the particle being at point $x_i$ after being observed is 
\begin{equation}
    p(x_i)= \frac{\lvert c_i\rvert^2}{\lvert \ket{\psi}\rvert^2} = \frac{\lvert c_i\rvert^2}{\sum_{j}^{} \lvert c_j\rvert^2}
\end{equation}

\subsection{Spin and Entanglement}
Even if quantum particles are isolated by long distance, they can collaborate with one another or the effect of the movement of one particle can affect another one. This phenomenon is called quantum entanglement. 

Spin is basic property of quantum particles and has no classical analogue. In 1922 the Stern-Gerlach Experiment \cite{yanofsky2008quantum} showed that an electron in the presence of a magnetic field  behaves as though it were a charged spinning top and acts as a small magnet. 
\begin{figure}[]
     \centering
     \includegraphics[scale=0.5]{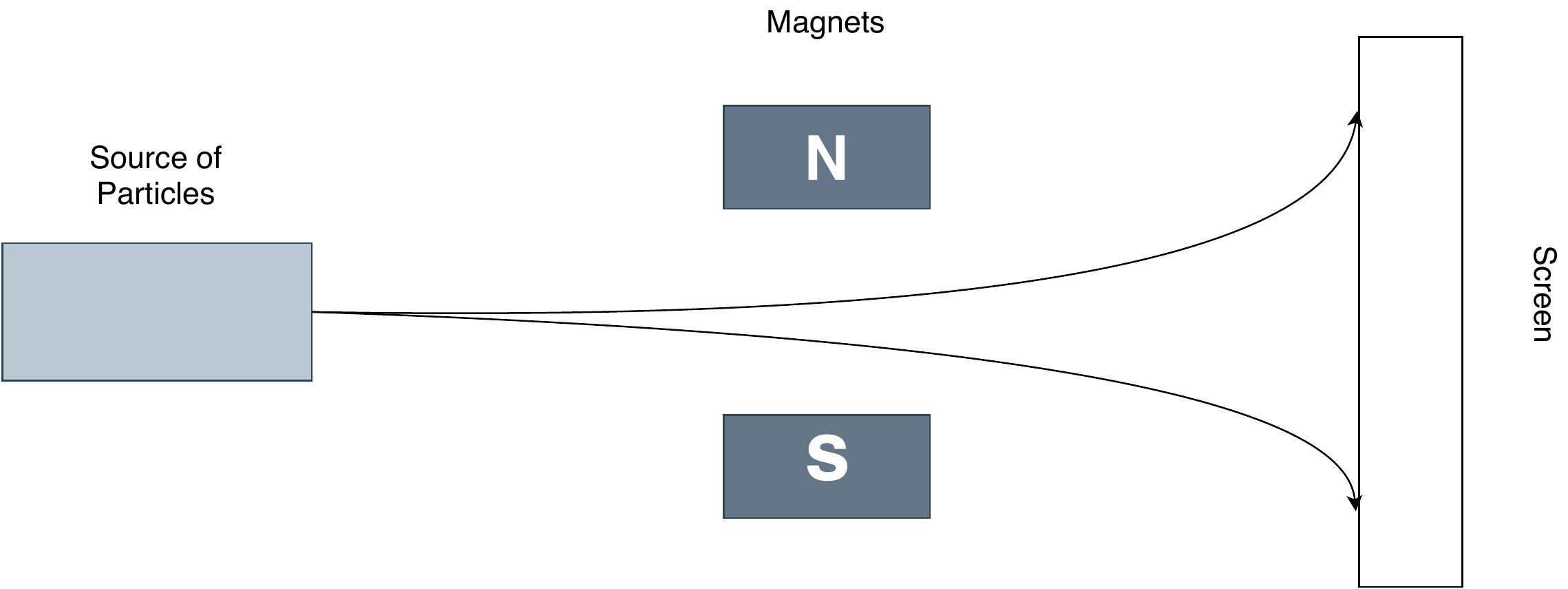}
     \caption{The Stern-Gerlach experiment\cite{yanofsky2008quantum}.}
     \label{fig:spin}
\end{figure}
As shown in Figure \ref{fig:spin} the source  shoots beams of electrons through a magnetic field which is nonhomogeneous. The electrons are supposed to hit  vertically directly into the screen. Bur they get splitted by the magnetic field into two opposite direction with opposite spin. Some electrons spin one way and some spin the opposite way. Now the electrons do not have an internal magnetic structure, they are just charged particles. From this experiment it is shown that the spinning particle when being measured in a specific direction can only be found either clockwise or anticlockwise state. For each of this given directions there exists two basic spin states. For example, for vertical axis these states are spin up $\ket{\uparrow}$ and spin down $\ket{\downarrow}$.  So the generate state becomes the superposition of of up and down i.e. 
\begin{equation}
    \ket{\psi} = c_0\ket{\uparrow} + c_1\ket{\downarrow}
\end{equation}
where $c_0$ and $c_1$ are respectively the amplitudes of finding the particle in the up state and down state.

Now that we have discussed about ``spin", if one particle of a pair is chosen to be in spin up state while being measured, at that point this choice is conveyed to the next connected particle that now acquires the opposite state spin ``down". Thus quantum entanglement allows far-off qubits to interact immediately with each other.

\subsection{Quantum Logic Gates}
In classical computing we have logical gates such as AND, OR, NOT gates which work on bits. But in the quantum world, only reversible gates work as quantum logical gates. These gates are represented by Unitary matrices and they work on qubits\cite{yanofsky2008quantum}. Operations that are not measurements such as NOT gate and Identity gates are reversible. OR gate is not reversible because given an output of $\ket{1}$ from OR, it cannot be determined if the input was $\ket{01}$, $\ket{10}$, $\ket{11}$. So the input cannot be determined from the output of the OR gate and hence OR is not a reversible gate.\\
The significance of reversible gates is that erasing information from classical computers causes energy loss and produces heat whereas quantum computing is reversible and does not erase information. Hence quantum computers do not lose energy. \\
Some significant quantum logic gates are:
\subsubsection{Controlled-NOT Gate}
Controlled-NOT gate has two inputs and two outputs. As showed in the circuit below, the top input called a control bit determines what the output will be. The bottom output of $\ket{y}$ will be the same as input if $\ket{x} = \ket{0}$. The output will be opposite if $\ket{x}=1$. 
\begin{center}
\begin{tikzcd}
\qw & \push{\ket{x}}  & \ctrl{1} & \push{\ket{x}} & \qw\\
\qw & \push{\ket{y}}  & \targ{} & \push{\ket{x\oplus y}} & \qw
\end{tikzcd}
\end{center}

From the circuit, the controlled-NOT gate takes $\ket{x,y}$ to $\ket{x,x\oplus y}$. The $\oplus$ notation indicates the binary exclusive OR operation. The matrix representing the controlled-NOT gate is \\
\[
\begin{blockarray}{ccccccc}
&&&00 & 01 & 10 & 11 \\
\begin{block}{ccc(cccc)}
  00 &&& 1 &0&0&0 \\
  01 &&& 0&1&0&0 \\
  10 &&& 0&0&1& 0 \\
  11 &&& 0&0&0&1 \\
\end{block}
\end{blockarray}
 \]

\subsubsection{Toffoli Gate}
Toffoli gate has two control bits. It is similar to controlled-NOT gate. The bottom bit flips only if both of the top bits are in state $\ket{1}$. Toffoli gate is also its own inverse. The logic circuit for toffoli gate is given below
\begin{center}
\begin{tikzcd}
\qw &\qw & \push{\ket{x}}&\qw  & \ctrl{1} &\qw &\push{\ket{x}} & \qw\\
\qw &\qw & \push{\ket{y}} &\qw & \ctrl{1} &\qw & \push{\ket{y}} & \qw\\
\qw &\qw & \push{\ket{z}} &\qw & \targ{} & \qw &\push{\ket{z\oplus (x\wedge y)}} & \qw
\end{tikzcd}
\end{center}

\subsubsection{Fredkin Gate}
Fredkin gate is another interesting gate. Fredkin gate has three inputs and three outputs which is similar to Toffoli gate. But it has only one control input. If $\ket{x}$,$\ket{y}$ and $\ket{z}$ are inputs then,$\ket{x\prime} = \ket{x}$. Now, if $\ket{s} = \ket{(y\oplus z)\wedge x}$. Then,$ \ket{y\prime} = \ket{y\oplus s}$ and $ \ket{z\prime} = \ket{z\oplus s}$. \\
In brief, $\ket{0,y,z} \longmapsto \ket{0,y,z}$ and  $\ket{1,y,z} \longmapsto \ket{1,z,y}$. Likewise Toffoli gate, Fredkin gate is also its own inverse.  The logic circuit for Fredkin gate is depicted below.
\begin{center}
\begin{tikzcd}
\qw &\qw & \push{\ket{x}}&\qw  & \ctrl{2} &\qw &\push{\ket{x}} & \qw&\qw\\
\qw &\qw & \push{\ket{y}} &\qw & \swap{} &\qw & \push{\ket{y\prime}} & \qw&\qw\\
\qw &\qw & \push{\ket{z}} &\qw & \swap{} & \qw &\push{\ket{z\prime }} & \qw&\qw
\end{tikzcd}
\end{center}

\subsubsection{Pauli Gates}
There are three Pauli matrices of $2\times 2$ dimension, which are also hermitian and unitary. The matrices are-
$X = \left[\begin{matrix}
0&1\\1&0
\end{matrix}\right]$ , $y = \left[\begin{matrix}
0&-i\\i&0
\end{matrix}\right]$ and $z = \left[\begin{matrix}
1&0\\0&-1
\end{matrix}\right]$.\\

There are also other quantum gates such as \textbf{square root of NOT} and is represented by $\sqrt{NOT} = \frac{1}{\sqrt{2}}\left[\begin{matrix} 1&-1\\1&1 \end{matrix}\right]$. But $\sqrt{NOT}$ gate is not its own inverse.

\subsection{Representative Quantum Algorithms}
\subsubsection{Quantum Fourier Transform}
The Quantum Fourier Transform which is the quantum analogue of the Discrete Fourier Transform is at the core of numerous quantum algorithms.  Discrete Fourier Transform is a square invertible matrix D of dimension N whereas QFT is defined by mapping each computational basis state $\ket{i}$ to a new quantum state $\ket{f_i} = \sqrt{\frac{1}{N}} \sum_{k=0}^{N-1}exp(j\frac{2\pi i k}{N})\ket{k}$ \cite{dervovic2018quantum}. 
\subsubsection{Grover's Algorithm}
Grover's Algorithm enables to find a special record in a randomly ordered database of $N$ records\cite{grover1996fast}. While a classical algorithm surely takes $\mathcal{O}(N)$ steps to find a specific record in that database, Grover's algorithm $\mathcal{O}(\sqrt{N})$ steps to find it.  
\subsubsection{Shor's Algorithm}
\begin{figure}[]
     \centering
     \includegraphics[scale=4.5]{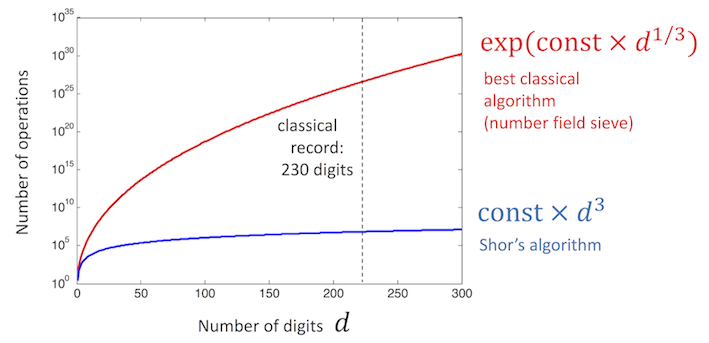}
     \caption{Classical vs quantum factoring algorithms\cite{IBM2017shor}.}
     \label{fig:clvsqu}
\end{figure}

Any integer number can be uniquely represented by the product of primes. But finding prime factors in classical computing is not an easy task. In fact, it is one of the hardest problems and so the security of our online transactions are based on the fact that, factoring an integer with thousands of digits is noting but impossible. This belief has been challenged by Peter Shor \cite{shor1994algorithms} when he proposed a new quantum method for prime factoring. His solution took $\mathcal{O}((n^3 \log n))$ time whereas the most efficient known algorithm take $\mathcal{O}(exp(\sqrt[3]{\frac{64}{9}n(\log n)^2}))$ \cite{pomerance2008tale} time. This figure \ref{fig:clvsqu} will show how dramatic the change has been made using the quantum computer. 
\subsection{Quantum Machine Learning Algorithms}
\subsubsection{The Harrow-Hassidim-Lloyd (HHL) Algorithm}
\label{hhl}
In this section, we discuss  about the Harrow-Hassidim-Lloyd (HHL) algorithm, a quantum algorithm for solving the system of linear equations\cite{harrow2009quantum}. In this paper, the authors first described why it is necessary to find a quantum solution for solving the system of linear equations efficiently. It is because the classical algorithm takes polynomial time that requires at least $N$ steps to solve the $N$ linear equations.  Then the authors described the problem and divided it into sections. The problem is to find $\bar{x}$ from the system $A\bar{x} = \bar{b}$ where $A$ is an $n\times n$ real Hermitian matrix and $\bar{b}$ is a vector. An $n \times n$ matrix is said to be a Hermitian matrix if $A^{\dag} = (\overline{A})^T = \overline{A^T} = A$ i.e. $A[j,k] = \overline{A[k,j]}$ \cite{yanofsky2008quantum}.  Classically $n^2$ steps would be necessary for analyzing all the entries of $A$ and $n$ steps would be needed even to  write out the solution. HHL assures to solve the system of $n$ equations  with complexity $O(\log N )$. The pseudo-code for the algorithm is reproduced  in Algorithm \ref{hhl} from \cite{dervovic2018quantum}. \\
\noindent\begin{minipage}{\textwidth}
\renewcommand\footnoterule{}    
\begin{algorithm}[H]
\caption{HHL Algorithm }
\label{hhl}
\KwInput{ State vector $\ket{b}$, matrix $A$ with oracle access to its elements. Parameters $t_{0} = O(\kappa/\epsilon)$, $T = \widetilde{\mathcal{O}}(\log(N)s^2t_0 )$, $\epsilon$ is desired precision.}
$\Lambda_{HHL}(\ket{b},A,t_0,T,\epsilon)$\{
\begin{enumerate}
    \item Prepare the input state $\ket{\Psi_0}^{C} \otimes \ket{b}^I$ , where $\ket{\Psi_0} = \sqrt{\frac{2}{T}}\sum\nolimits_{\tau = 0}^{T-1} \sin{\frac{\pi(\tau+\frac{1}{2})}{T}}\ket{\tau}^{C} $.
    \item Apply the conditional Hamiltonian evolution $\sum\nolimits_{\tau = 0}^{T-1}\bra{\tau}\ket{\tau}^{C} \otimes e^{\frac{iA\tau t_0}{T}}$ to the input.
    \item Apply the quantum Fourier transform to the register $C$, denoting the new basis states $\ket{k}$, for $k \in {0,....T-1}$. Define $\widetilde{\lambda} = 2\pi k/t_0$.
    \item Append an ancilla register, $S$, and apply a controlled rotation on $S$ with $C$ as the control, mapping states $\ket{\widetilde{\lambda}} \longmapsto \ket{h(\widetilde{\lambda})}$, with $\ket{h(\widetilde{\lambda})}$\footnote{$\ket{h(\widetilde{\lambda})}$ is a state which comes after the controlled rotation of register S described in \cite{dervovic2018quantum} in details}. 
    \item Uncompute garbage in the register $C$.
    \item Measure the register $S$.
    \item \textbf{if}( result = `well' ) { \textbf{return} register $I$ }\\
\textbf{else}{ goto step $1$. }
\end{enumerate}

\}\\
Perform $O(\kappa)$ rounds of amplitude amplification on $\Lambda_{HHL}(\ket{b},A,t_0,T,\epsilon)$.\\
\KwOutput{State $\ket{\widetilde{x}}$ such that $||\ket{\widetilde{x}} - \ket{x}|| _2\leq \epsilon$ .} 
\end{algorithm}
\end{minipage}\\

First, the matrix $A$ needs to be invertible so that both sides of $A\bar{x} = \bar{b}$ can be multiplied by $A\textsuperscript{-1}$. The authors mentioned about a crucial factor $\kappa$ defining the ratio between $A's$ largest and smallest eigenvalues called condition number of $A$. If this  condition number grows, the solution becomes less static as $A$ becomes less invertible. In this situation the authors named the matrix to be ``ill-conditioned". But if the complexity of $\kappa$ becomes $\text{poly}( \log(n))$, the algorithm achieves an exponential speedup. \\
The hermitian matrix $A$ needs to be transformed into a unitary operator $e\textsuperscript{iAt}$ for different values of $t$. For this $A$ needs to be $s$ sparse i.e. $s$ non-zero entries per row so that these entries can be computed in time complexity $O(s)$. Under these circumstances, $e\textsuperscript{iAt}$ can be simulated in time $O(\log(N)s^2t)$ according to \cite{berry2007efficient}. But if $A$ is not hermitian, a reduction is needed to be applied. \\
Then $b$ needs to be prepared. The HHL algorithm interprets $b$ as a quantum state $\ket{b} = \sum_{j=1}^{n} b_i\ket{i}$. The authors in s \cite{harrow2009quantum} mentioned that the HHL could be a function in a larger quantum algorithm in which $\ket{b}$ would be produced by some other parts. In the next step, $\ket{b}$ is disintegrated using phase estimation \cite{luis1996optimum, cleve1998quantum, buvzek1999optimal} into the eigenvector basis. After this, the conditional Hamiltonian simulation \cite{berry2007efficient,  childs2010relationship} is used to apply the unitary operator to $\ket{b}$ for different $t$.  \\
Coles et al. in \cite{coles2018quantum} noted three sets of qubits named a single ancilla qubit,  an $n$ qubit register and a memory of $O(\log(n))$. In quantum world, a qubit can be in one state and in another state simultaneously i.e. a qubit can represent true and false or state $ \ket{0} $ and  $ \ket{1} $ at the same time \cite{yanofsky2008quantum}. The authors in \cite{coles2018quantum} divided the HHL algorithm into three parts: (a) phase estimation using unitary transformation, mapping the eigenvalues into the register in binary form with precision up to $n$ bits, (b) rotation of the ancilla qubits for each eigenvalue, performed through controlled rotation on the ancilla qubit , which yields 
\begin{equation}
\label{eqn : 1}
    \sum_{j=1}^{n}\beta_j\left(\sqrt{1- \frac{C^2}{(\lambda_j)^2}}\ket{0}+ \frac{C}{\lambda_j}\ket{1}\right)\ket{\lambda_j}\textsubscript{register}\ket{u_j}_\textsubscript{memory} 
\end{equation} \cite{coles2018quantum}
where $C$ is a normalizing constant, ${\ket{u_j}}$ and ${\lambda_j}$ are the eigenbasis and eigenvalues of A respectively and $\ket{b} = \sum_{j=1}^{n}\beta_j\ket{u_j}$, 
(c)  by reversing the first step i.e. undoing the phase estimation  
\begin{equation}
    \sum_{j=1}^{n}\beta_j\left(\sqrt{1- \frac{C^2}{(\lambda_j)^2}}\ket{0}+ \frac{C}{\lambda_j}\ket{1}\right)\ket{u_j} 
\end{equation} \cite{harrow2009quantum} is obtained,   which further gives 
\begin{equation}
    \ket{x} \approx \sum_{j=1}^{n} C\left( \frac{\beta_j}{\lambda_j}\right)\ket{u_j} 
\end{equation} 
\cite{coles2018quantum} by selecting $\ket{1}$ state. \\
Coles et al. \cite{coles2018quantum} applied the HHL algorithm on a $2\times 2$ system using $4$ qubits : one ancilla qubit, one memory, two register qubits . In their experiment, the eigenvalues of $A$ were $\lambda_1 = 1$, $\lambda_2 = 2$ and the eigenvectors were $\frac{1}{\sqrt{2}} \left(\begin{matrix}1\\-1\end{matrix}\right) $, 
$\frac{1}{\sqrt{2}} \left(\begin{matrix}1\\-1\end{matrix}\right) $ or $\ket{-}$ , $\ket{+}$ respectively. They used controlled $U$ rotation with $\theta = \pi$ for $\lambda_1$ and $\theta = \pi/3$ for $\lambda_2$ and $C=1$ for Eq. \ref{eqn : 1}. For $b$ they used three cases: $ \left(\begin{matrix}1\\0\end{matrix}\right) $,  $\frac{1}{\sqrt{2}} \left(\begin{matrix}1\\-1\end{matrix}\right) $,  
$\frac{1}{\sqrt{2}} \left(\begin{matrix}1\\-1\end{matrix}\right) $. The authors depicted a comparison between the theoretical values and the results that were found in local simulator. They also showed a comparison  between the simulator result and the results from ibmqx4 machine with Z measurement on the equivalent composer circuit.  \\
In \cite{aaronson2015read} the author mentioned four caveats of the algorithms. They are :  
\begin{enumerate}
    \item The vector $b$ is required to be loaded into the memory very fast so that a quantum state $\ket{b} = \sum_{i=1}^{n} b_i\ket{i}$ can be prepared.
    \item The quantum computer needs to be eligible for  applying unitary transformation efficiently. For this purpose  $A$ should be $s$ sparse which is discussed before. 
    \item Another important factor is the matrix $A$ requires to be robustly invertible i.e. ``well-conditioned". To be more exact, $\kappa$ which is the ratio of $\lambda\textsubscript{max}$ and $\lambda\textsubscript{min}$. If $\kappa$ grows exponentially, the amount of time needed by HHL also grows linearly which gets rid of exponential  speedup. 
    \item The output of $x$ is a quantum state $\ket{x}$ of $\log_2(n)$ qubits.  Learning the value of any specific entry $x_i$ will, by and large, demand recurring the algorithm roughly n times which can terminate the exponential speedup. 
\end{enumerate}

These four caveats are so significant that they were further  reevaluated  in \cite{shao2018reconsider}.  Additionally, a few more caveats about HHL algorithm were discussed  by the author such as the singular values of $A$ locate between $1/\kappa$ and $1$ and this caveat derives from quantum phase estimation to estimate the eigenvalues of Hermitian matrix. According to the author, the eigenvalues should be compressed between $0$ and $2\pi$. The author also briefly discussed  about the solution of the four caveats mentioned earlier. Some quantum machine learning algorithms related to the HHL algorithm were elaborated in this paper.  

\subsubsection{Quantum Support Vector Machine}
\label{qsvm}
Support Vector Machine (SVM), one of the most popular algorithms in modern machine learning, was introduced by Vapnik in 1992 \cite{marsland2014machine}. It is a supervised machine learning algorithm that classifies vectors in a feature space into one of two  binary classes, given reasonably sized datasets. SVM does not work well on enormously large datasets as it becomes computationally expensive. \\
Given $M$ data points of $(\overrightarrow{x_j},y_j):j=1,2,....,M$, SVM classifies a vector into one of the two classes, where $\overrightarrow{x}$ is an $N$-dimensional vector in the feature space and $y_j = +1$ or $-1$ which is the label of the data\cite{coles2018quantum}. Having a weight vector $\overrightarrow{w}$, an input vector $\overrightarrow{x}$ and a bias weight $b$, SVM finds the hyperplane with maximum margin. The hyperplane $\overrightarrow{w}+\overrightarrow{x} = b$ divides the data points into  two classes so that $\overrightarrow{w}+\overrightarrow{x_j} \geq 1$ and $\overrightarrow{w}+\overrightarrow{x_j} \leq 1$ when $y_j = +1 $ and $y_j = -1$ respectively. \\
Least Squares SVM (LS-SVM) , a  variant of SVM formulated in\cite{suykens1999least} and also mentioned in \cite{rebentrost2014quantum, coles2018quantum}, estimates the hyperplane for  obtaining the procedure of SVM by solving the  system of linear equation in \ref{eqn : svm}:
\begin{equation}
    \label{eqn : svm}
        \left[\begin{matrix}0&\overrightarrow{1}^T\\\overrightarrow{1}&K+\gamma^{-1}1 \end{matrix}\right] \left[\begin{matrix}b\\\overrightarrow{a} \end{matrix}\right] = \left[\begin{matrix}0\\\overrightarrow{y} \end{matrix}\right]
\end{equation}.

Here $K$, the $M\times M$ dimensional matrix , is called the kernel matrix, $\gamma$ is the tuning parameter and $\overrightarrow{a}$ forms the vector $\overrightarrow{w}$ where $\overrightarrow{w} = \sum\nolimits_{j=1}^{M} a_j\overrightarrow{x_j} $.  Different types of kernels are in there but linear kernel ($K_{ij} = \overrightarrow{x_i}\cdot \overrightarrow{x_j}$) is used in Quantum Support Vector Machine\cite{rebentrost2014quantum}. The classical complexity of least square support vector machine is $O(M^2(M+N))$ and the complexity of Quantum Support Vector Machine is $O\log(NM)$.      \\

The pseudo-code of  quantum support vector machine for solving linear equations is reproduced in \ref{qsvmalgo} from \cite{coles2018quantum}.

\noindent\begin{minipage}{\textwidth}
\begin{algorithm}[H]
\caption{Quantum Support Vector Machine Algorithm }
\label{qsvmalgo}
\KwInput{
\begin{itemize}
    \item Training data set ${(\overrightarrow{x_j},y_j): j= 1,2,....,M}$.
    \item A query data $\overrightarrow{x}$.
\end{itemize} 
}

\KwOutput{Classification of $\overrightarrow{x}:$ $+1$ or $-1$} 

\textbf{Procedure:}
\begin{enumerate}
    \item Calculate kernel matrix $K_{ij} = \overrightarrow{x_i}\cdot \overrightarrow{x_j}$ using quantum inner product algorithm\cite{lloyd2013quantum}.
    \item Solve linear equation \ref{eqn : svm} and find $\ket{b.\overrightarrow{\alpha}}$ using a quantum algorithm for solving linear equations \cite{giovannetti2008quantum} (training step).
    \item Perform classification of the query data $\overrightarrow{x}$ against the training results $\ket{b,\overrightarrow{\alpha}}$ using a quantum algorithm \cite{rebentrost2014quantum}.
\end{enumerate}
\end{algorithm}
\end{minipage}

%% file: tex/foundation.tex
\label{foundation}


As integrating machine learning with quantum computing is comparatively a new concept, we are going to present some of the previous works on this field all together. It will help readers to have an overview of the recent works done by researchers on this field. We will also focus on the researches on the basic quantum computing too. We have reviewed the following papers and blogs.

\begin{itemize}
    \item An introduction to quantum machine learning
    \item Read the fine print
    \item Preconditioned quantum linear system algorithm
    \item Making  a  Neural  Network,  Quantum
    \item Quantum convolutional neural networks
    \item Generative model benchmarks for superconducting qubits
    \item A generative modeling approach for benchmarking and training shallow quantum circuits
    \item Quantum algorithm implementations for beginners
    \item A survey of quantum learning theory
\end{itemize}

\par In this paper\cite{schuld2015introduction}, the authors gave a systematic overview of  quantum machine learning. The authors tried to present the methods of quantum machine learning as well as their computational details in an accessible way. They also discussed the future potentials of the quantum machine learning. As this contribution states, the volume of the globally stored data is growing by around 20\% every year. So, the necessity of learning algorithms are increasing. The need of interpreting the data based on experience is rising. The authors focus on the process of exploitation of the potentials of quantum computing in order to optimise classical machine learning algorithms. From this paper, we come to know that, the laws of quantum mechanics restrict us to have an access to the information stored in quantum systems and coming up with quantum algorithms that can outperform their classical counterparts are very difficult. However, the authors mentioned that the toolbox of quantum algorithms is by now fairly established and also contains a number of impressive examples that speed up the best known classical methods. Their work also sited other researchers works. The authors referred to some other contributions where other scientists are trying to develop quantum algorithms that can recognise patterns. 
\par The main focus of this contribution is to recognise patterns with quantum machine learning algorithms. Firstly, the authors described the concepts of classical and quantum learning in brief. Then they divided the paper in seven sections and in each section they described a standard method of machine learning. The methods they described are namely k-nearest neighbour methods, support vector machines, k- means clustering, neural networks, decision trees, Bayesian theory and hidden Markov models. 
\par In the very next section the authors took a quick look into the classical and quantum learning theory. They mentioned the fact that the theory of machine learning is an interdisciplinary subject of both artificial intelligence and statistics. In theory of machine learning the term learning is divided into three categories namely supervised learning, unsupervised learning and reinforcement learning. The authors discussed these three types of learning in short. The authors claimed that, whatever learning method is chosen, optimal machine learning algorithms run with minimum resources and have a minimum error rate related to the task. They think that, challenge is to find the optimal parameter and values that can lead us to a good solution. Otherwise, we can come up with schemes that reduce the complexity   of the algorithm. The authors pointed out the fact that, quantum machines can help us on this point. 
\par The next section contains the brief story of quantum machine learning. This term refers to the manipulation of quantum systems to process information. The authors claimed that the characteristics of quantum system which includes the superposition of quantum particles can help us in  speeding up the computations in terms of complexity. As our previous sections describe, this paper also has the description of qubits and the superposition of the qubits. This paper has shown some of the mathematical concepts of the quantum system. The authors briefly discussed the unitary quantum gates also in this paragraph. Here, they have shown the basic XOR gate working on the quantum state. In quantum machine learning, quantum algorithms are developed to solve typical problems of machine learning using the efficiency of quantum computing. 
\par In the next section, they discussed the quantum versions of machine learning algorithms. They discussed the previously mentioned algorithms in terms of a quantum perspective. There are many technical details discussed in this paper as well as in the following papers which are reviewed here. Here, for k-nearest neighbours methods, support vector machines and k-means clustering, the authors were mainly concerned to find the efficient calculations of classical distances on a potential quantum computer. But they did not do the same for all other algorithms. Two probabilistic methods, Bayesian theory and hidden Markov models, were used to find an analogy in the formalism of open quantum systems. Again, neural networks and decision trees were used to have first explorations of quantum models. 
\par This paper was our main concern for further progression in this quantum machine learning field as this is kind of an introductory paper. This paper gave an overview of existing ideas and approaches to quantum machine learning. The authors stated that a quantum theory of learning is yet impressive. Although while working on quantum machine learning algorithms, only very few contributions actually answer the question of how the main feature of machine learning, the learning process, can actually be simulated in quantum systems. The authors suggested to investigate the different approaches to quantum computing for this purpose. The authors mentioned the contributors\cite{bisio2010optimal}, who have already  done some of the investigations on the ideas in that directions. Last of all, they are hopeful that even there is still a lot of work to do and quantum machine learning remains a very promising emerging field of research with many potential applications and a great theoretical variety. 
\par Here in this \cite{aaronson2015read} paper, the author has discussed the HHL \cite{harrow2009quantum} algorithm. We come to know from this paper that HHL algorithm can solve the system of linear equations. The goal of HHL is to solve the system, \\
\begin{equation}
    Ax = b
\end{equation}  
HHL can solve this equation in logarithmic time, that is in $ \log(n) $ time. Here, $ n $  depicts the number of unknowns or the number of equations in the linear system. Classically, this complexity seems unreal. As, we need at least $ n^2 $ of steps to examine the entries of $ A $ and more $ n $ steps will be needed to print the solution matrix $ x $. The author summarized the whole concept like this, HHL does not exactly solve the system of linear equations in $ \log(n) $ time. It prepares the quantum superposition of the form $ \ket{x} $. Here, $ x $ is the solution vector to the system of linear equation.
\par The author made a statement that, we can use HHL as a template for other quantum algorithms. Here, in this \cite{clader2013preconditioned} Clader et al. claimed that we can use HHL to speed up the calculations of electromagnetic scattering cross-sections, for systems involving smooth geometric figures in 3-dimensional space. In the next section in this paper, the author has shown the mathematical calculation of the fact that HHL algorithm can solve the equations of linear system in logarithmic time. In this thesis work we have already implemented the HHL algorithm on a quantum machine. We have discussed that in later section. 
\par After learning about the quantum HHL algorithm we shifted our focus on a blog post, which contains contents about neural networks in quantum machine learning. This article  \cite{bromley2018making} is all about quantum neural network. In this blog post, the author discussed about the process of implementing the quantum neural network. This will create a good opportunity for computer science. In a figure used in the blog post the author has shown that we can store any configuration of 4 neurons in only 2 qubits. This unlocks our access to a huge number of quantum algorithms that can be implemented to boost processing the network. For this purpose, finding the perfect algorithm to advance the performance is granted as the first task by the author. 
\par The author mentioned about the Hopfield Network which can be used as a addressable memory system. Conventional way of using the Hopfield network is to keep picking neurons at random and updating them by considering the connected neurons, along with their weights. Adding the embedding system into qubits makes the quantum HHL algorithm. This algorithm can process the Hopfield Network. This helps to have an advantage in storage capacity. Researchers have highlighted in particular the application of the Hopfield network within genetics as a recognizer of infectious diseases also. The author shortly described the next tasks of this field. The author mentioned about the eagerness to uncover improvements to the Hopfield network through quantum mechanics. The speed of reading and writing the data in a quantum machine is still needed to be addressed as the author mentioned. Many experimental teams are working on the technological development like chip design and implementing photonic quantum processors. 
\par Next we are going to review this \cite{cong2018quantum} paper, with the title "Quantum Convolutional Neural Networks". Being motivated by the progress in realization of quantum information processors the authors in this paper introduced and analyzed a novel quantum machine learning (QCNN) model which is basically inspired by convolutional neural networks. They claimed that QCNN circuits combine the multi-scale entanglement re-normalization ansatz and quantum error correction to mimic renormalization-group flow. In this contribution, they proposed a model for QCNN circuit. The input to the circuit model is an quantum state $ \rho_{in} $. 
\par Later, experimental considerations have been discussed. This contribution claims that the QCNN architecture they provided can be efficiently implemented on several state of the art experimental platforms. The key ingredients for realizing QCNNs include the efficient preparation of quantum many-body input states, as author stated, the application of two-qubit gates at various length scales, and projective measurements. These capabilities have already been implemented in multiple programmable quantum simulators consisting of $ N \geq 50 $ qubits, which are actually based on trapped neutral atoms and ions, or superconducting qubits. Authors are expecting that, the limited depth of their circuit which is approximately $ log(N ) $ may allow its implementation even in the near term experiments with relatively short coherence time. 
\par Next, they discussed about the further scopes for QCNN to work with. As this contribution only presented the QCNN circuit structure for recognizing 1D phases, which is straightforward to generalize the model to higher dimensions. They are hoping that, studying QCNNs in two and higher dimensions could potentially help identify nonlocal order parameters for lesser-understood phases.
\par Let us move onto another paper \cite{hamilton2019generative}. Here, in this paper, the authors discussed about three performance measures for Quantum Machine Learning. They are:  Kullbeck-Leiber divergence which compares  the target and output distribution , F1 score which measures how well a circuit each state and qBAS22 which is a modified F1 score. Using these metrics the authors measured how well each circuit can learn the target distribution in the absence of noise. They used different different entangling layers (1-2) with different CNOT gates (2-4) on noiseless qubits and showed the minimum divergence and maximum qBAS scores. 
\par The authors showed that the behaviour of Kullbeck-Leiber divergence is affected with presence of noise. The authors found smallest KL and largest qBAS on IBM Tokyo for  entangling layer 2 with 2 CNOT gates. Then the authors showed how QCBM (Quantum Circuit Born Machine) circuit performs on noisy qubits with parameters which are optimized with noisy qubits. Here noise means qubit decoherence, gate infidelity and measurement errors. The authors mentioned that goal of classical machine learning in general is to avoid over-fitting and to tune parameters properly where QML relies on  limiting number of training steps and rotational parameters. They discussed about the effect of noise and N\textsubscript{shots} (sampling) in details. At last they talked about future works for improving noise-resilience of circuit training and other areas for error mitigation for better estimation.
\par To clarify the topic of quantum machine learning we tried to cover more and more research works. As this topic is an advanced topic we spent more time on reading research papers. Now, we are going to present another research work \cite{benedetti2019generative}, which discusses the details on NISQ machines. As hybrid quantum-classical algorithms provides a way to use the noisy intermediate-scale(NISQ) quantum computers for practical purposes, the authors tried to expand the portfolio of these techniques and proposed a quantum circuit learning algorithm that can be used to assist the characterization of quantum devices and to train shallow circuits for generative tasks. The authors' approach can learn an optimal preparation for famous Greenberger-Horne-Zeilinger states which is also known as "cat states". In quantum information theory, a Greenberger-Horne-Zeilinger state is a special type of quantum entangled state which involves at least three particles. 
\par In this contribution, the authors claimed that their proposed method can efficiently prepare approximate representation of coherent thermal states, wave functions that encode Boltzmann probabilities in their amplitudes. The authors designed a hybrid quantum-classical framework called data-driven quantum circuit learning (DDQCL). The authors simulated quantum circuits using the QuTiP2 Python library and implemented the constraints dictated by the ion trap experimental setting. They worked with three synthetic data sets: zerotemperature ferromagnet, random thermal, and bars and stripes (BAS). In all of their numerical experiments the authors used 1000 data points sampled exactly from these distributions.
\par This time we put focus on the implementation of some quantum algorithms. This paper \cite{coles2018quantum} helped us having the knowledge of implementing the algorithms. Quantum computing uses some quantum-mechanical phenomenons, in particular, superposition, entanglement and quantum tunneling. This article describes the principles of quantum programming with straight forward algebra that makes understanding under laying quantum mechanics. The authors have explained 20 different quantum algorithms. Those algorithms are namely Grover's Algorithm, Bernstein-Vazirani Algorithm, Shor's Algorithm, Quantum Minimal Spanning Tree, Quantum Maximum Flow Analysis, Quantum Approximate Optimization Algorithm, Quantum Principal Component Analysis, Quantum Support Vector Machine, Quantum Simulation of the Schrodinger Equation, Quantum Simulation of the Transverse Ising Model, Quantum Partition Function, Quantum Tomography and some more. They have shown, how to implement those algorithms on IBM's quantum computers. They have also discussed about the results of those implementations with respect to differences of the simulator and the actual hardware runs. The authors also experimented with simplest quantum finite automata simulation to study whether quantum error correction (QEC) can improve computation accuracy. They have discovered that, implementation of the QEC does not improve the probability to interpret the final outcome correctly.
\par In this paper \cite{coles2018quantum} the authors have shown the mathematical implementation of those algorithms too. On a certain point they introduced the readers with the basic gates of the quantum system like Hadamard Gate (or Hadamard Matrix) \cite{wiki2019hadamard}. The authors classified the quantum algorithms in several groups like number-theory-based, oracle-based, and quantum simulation algorithms.
\par Now we are going to review this survey of quantum learning theory by Srinivasan Arunachalam and Ronald de Wolf \cite{arunachalam2017guest}. In this paper the authors survey quantum learning theory and the theoretical aspects of machine learning using quantum computers. They describe the main results known for three models of learning namely exact learning from membership queries, and Probably Approximately Correct (PAC) and agnostic learning from classical or quantum examples. At the beginning part, the authors described the history of Quantum Computing briefly, how this field of knowledge developed and how the Shor's efficient quantum algorithms for factoring integers and computing discrete logarithms gave momentum to this field. 
\par This contribution focuses on the theoretical aspects of quantum machine learning. They briefly discussed the introduction to quantum information system. Before jumping into learning algorithms they discussed about the basic quantum system including the Qubits, Bra-ket notations, multi-qubit states, quantum states, Hadamard Transformation etc. Then they moved into the query model analysis. After this the authors explored some details about Grover's algorithm and Fourier Sampling. 
\par In the next section, the authors jumped into the Learning models. Here they described the learning models they used in details. They used Exact Learning, PAC Learning and Agnostic Learning. They described the quantum model of those three learning as well as the classical model of them. As the PAC concerns a nice formalism for deciding how much data we need to collect in order for a given classifier to achieve a given probability of correct predictions on a given fraction of future test data. 
\par After describing all of the learning methods they started discussing the results of learning models. Besides, the gave a briefing on the complexity for each learning algorithm in terms of time complexity and memory complexity. Next they discussed the learning ability of quantum states. 
\par In our contribution, we have tried to implement some of those papers' methods and have presented some of the experiments and their results described in the next sections.

%% file: tex/ibm.tex
\label{ibm}
Nowadays anyone can use a Quantum Computer from anywhere in the world. Many famous companies are providing the opportunities to get the quantum experience online. \href{https://www.ibm.com/bd-en}{\textit{IBM}} is one of them. 

Quantum computers can be constructed with trapped ions. IonQ makes platform for trapped ion quantum computers. Google, IBM and some other companies produces the platform for superconducting quantum computers. Superconducting quantum computing is an implementation of quantum computer. As a quantum computer cannot lose energy and information for the sake of reverse computing, it has to be of zero resistant \cite{hui2019qc}. The IBM quantum computer is made of superconducting material and consists of superconducting  circuits. 

For all of our experiments, we have utilized the backends administered by IBM. Some of them are original quantum devices and some of them are classical devices which simulate the experiments as a quantum device will do. So, we provide some details about these backends. This section will be convenient for readers to understand the architecture of the devices on which we have run our experiments.

IBM provides nine backends on cloud for any user on the world. Each of them can be accessed from \url{https://quantum-computing.ibm.com/}. Table  \ref{tab:ibm} demonstrates a concise overview of each backends.

\begin{table}[]
\caption{An Overview of IBM Backends.}
\label{tab:ibm}
\begin{tabular}{|l|c|c|c|c|}
\hline
\multicolumn{1}{|c|}{\textbf{Gate Name}} &
  \textbf{\begin{tabular}[c]{@{}c@{}}Opened \\ Since\end{tabular}} &
  \textbf{\begin{tabular}[c]{@{}c@{}}No. of \\ Qubits\end{tabular}} &
  \textbf{Device Type} &
  \textbf{Basic Gates} \\ \hline
ibmq\_16\_melbourne v2.0.2        & Nov 06, 2018 & 15 & Real Quantum & u1,u2, u3,cx,id \\ \hline
ibmq\_london v1.0.0               & Sep 13, 2019 & 5  & Real Quantum & u1,u2, u3,cx,id \\ \hline
ibmq\_burlington v1.1.4           & Sep 13, 2019 & 5  & Real Quantum & u1,u2, u3,cx,id \\ \hline
ibmq\_essex v1.0.1                & Sep 13, 2019 & 5  & Real Quantum & u1,u2, u3,cx,id \\ \hline
ibmq\_ourense v1.0.1              & Jul 03, 2019 & 5  & Real Quantum & u1,u2, u3,cx,id \\ \hline
ibmq\_vigo v1.0.2                 & Jul 03, 2019 & 5  & Real Quantum & u1,u2, u3,cx,id \\ \hline
ibmq\_5\_yorktown - ibmqx2 v2.0.5 & Jan 24, 2017 & 5  & Real Quantum & u1,u2, u3,cx,id \\ \hline
ibmq\_armonk v1.1.0               & Oct 16, 2019 & 1  & Real Quantum & id,u1,u2, u3    \\ \hline
ibmq\_qasm\_simulator v0.1.547 &
  May 02, 2019 &
  32 &
  \begin{tabular}[c]{@{}c@{}}Quantum \\ Simulator\end{tabular} &
  \begin{tabular}[c]{@{}c@{}}u1,u2,u3,cx,cz,\\ id,x,y,z,h,s,sdg, \\ t,tdg, ccx,swap,\\  unitary,\\ initialize,kraus\end{tabular} \\ \hline
\end{tabular}
\end{table}

Brief description of the $9$ backends are given in the following.

\subsection{ibmq\_16\_melbourne}
\begin{figure}[H]
\centering
  \includegraphics[scale=0.5]{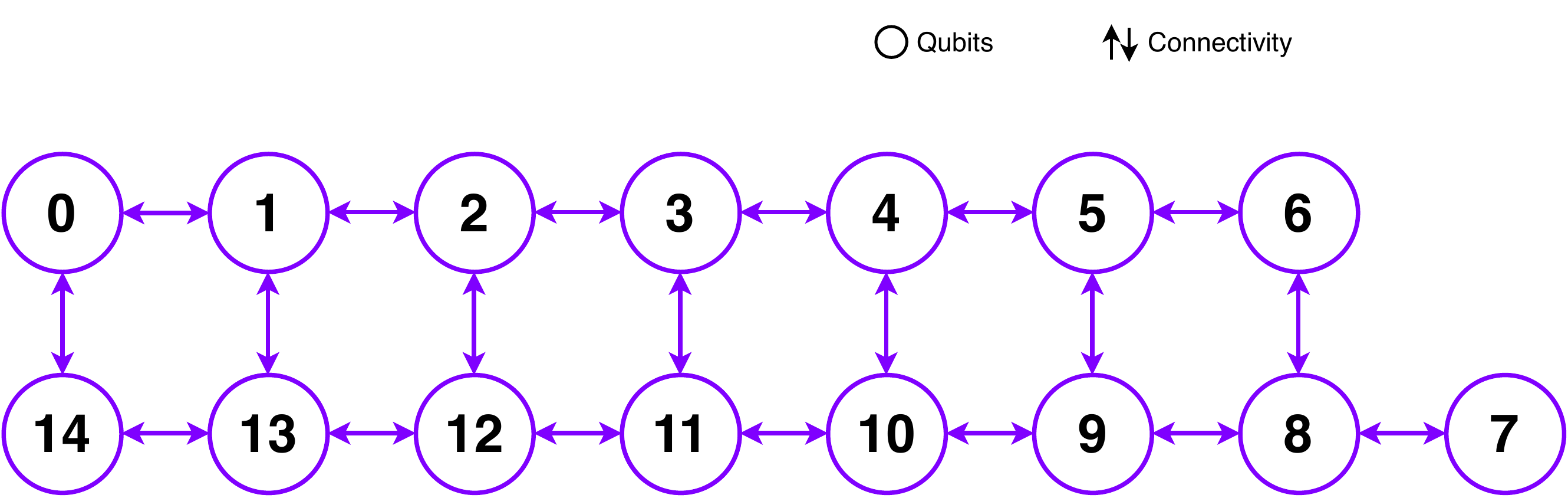}
  \caption{Hardware Architecture of ibmq\_16\_melbourne.}
  \label{fig:ibm16melbourne}
\end{figure}

This is a real quantum hardware. It has got $15$ qubits. It supports the basics gates- id, u1, u2, u3 and cx. Single-qubit U2 error rate is in between $3.644e-4$ and $2.706e-3$. CNOT error rate is in between $1.683e-2$ and $8.174e-2$. Figure \ref{fig:ibm16melbourne}   shows the hardware architecture of this machine using qubits and connections.

\subsection{ibmq\_london}
\begin{figure}[]
\centering
  \includegraphics[scale=0.5]{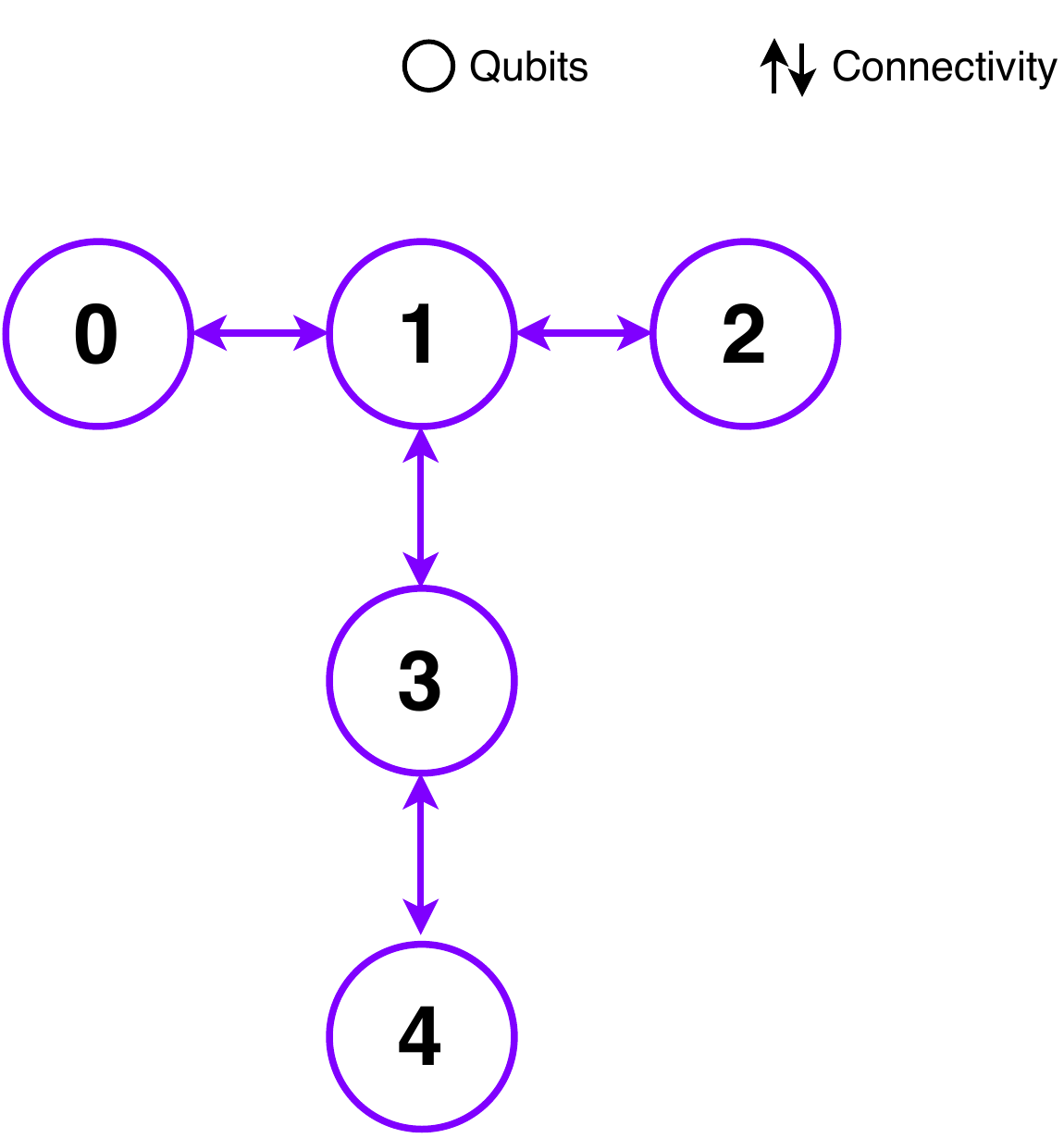}
  \caption{Hardware Architecture of ibmq\_london, ibmq\_burlington, ibmq\_essex, ibmq\_ourense, ibmq\_vigo. They look all the same but their error spectrum is different.}
  \label{fig:ibmlondon}
\end{figure}

This is again a real quantum hardware. It has taken only 5 qubits. It reinforces the basics gates- u1, u2, u3, cx and id. Single-qubit U2 error rate is in between $2.864e-4$ and $5.819e-4$. CNOT error estimate is in between $9.056e-3$ and $3.028e-2$. The architecture design is depicted in figure \ref{fig:ibmlondon}.
\subsection{ibmq\_burlington}
This real quantum hardware holds 5 qubits. It supports the basics gates- u1, u2, u3, cx and id. Single-qubit U2 error rate is in between $4.028e-4$ and $7.133e-4$. CNOT error percentage is in between $1.081e-2$ and $2.210e-2$. The design of its architecture is as same as \ref{fig:ibmlondon} with a different error rate.

\subsection{ibmq\_essex}
This real quantum hardware consists of 5 qubits. It reinforces the basics gates- u1, u2, u3, cx and id. Single-qubit U2 error rate is in between $3.405e-4$ and $8.875e-4$. CNOT error score is in between $9.416e-3$ and $1.533e-2$. Figure \ref{fig:ibmlondon} is the design of this machine but with a different error rate.

\subsection{ibmq\_ourense}
This is still a real quantum hardware. It has got only 5 qubits. It supports the basics gates- u1, u2, u3, cx and id. Single-qubit U2 error rate is in between $1.988e-4$ and $8.408e-4$. CNOT error rate is in between $6.319e-3$ and $1.766e-2$. Figure \ref{fig:ibmlondon} is also the design of  ibmq\_ourense. 

\subsection{ibmq\_vigo}
This real quantum hardware consists of 5 qubits. It supports the basics gates- u1, u2, u3, cx and id. Single-qubit U2 error rate is in between $3.660e-4$ and $1.062e-3$. CNOT error rate is in between $7.586e-3$ and $1.414e-2$. The  architechture of ibmq\_vigo is also same as four other machines shown in \ref{fig:ibmlondon} and with a different level of spectrum.

\subsection{ibmq\_5\_yorktown - ibmqx2}
\begin{figure}[]
\centering
  \includegraphics[scale=0.5]{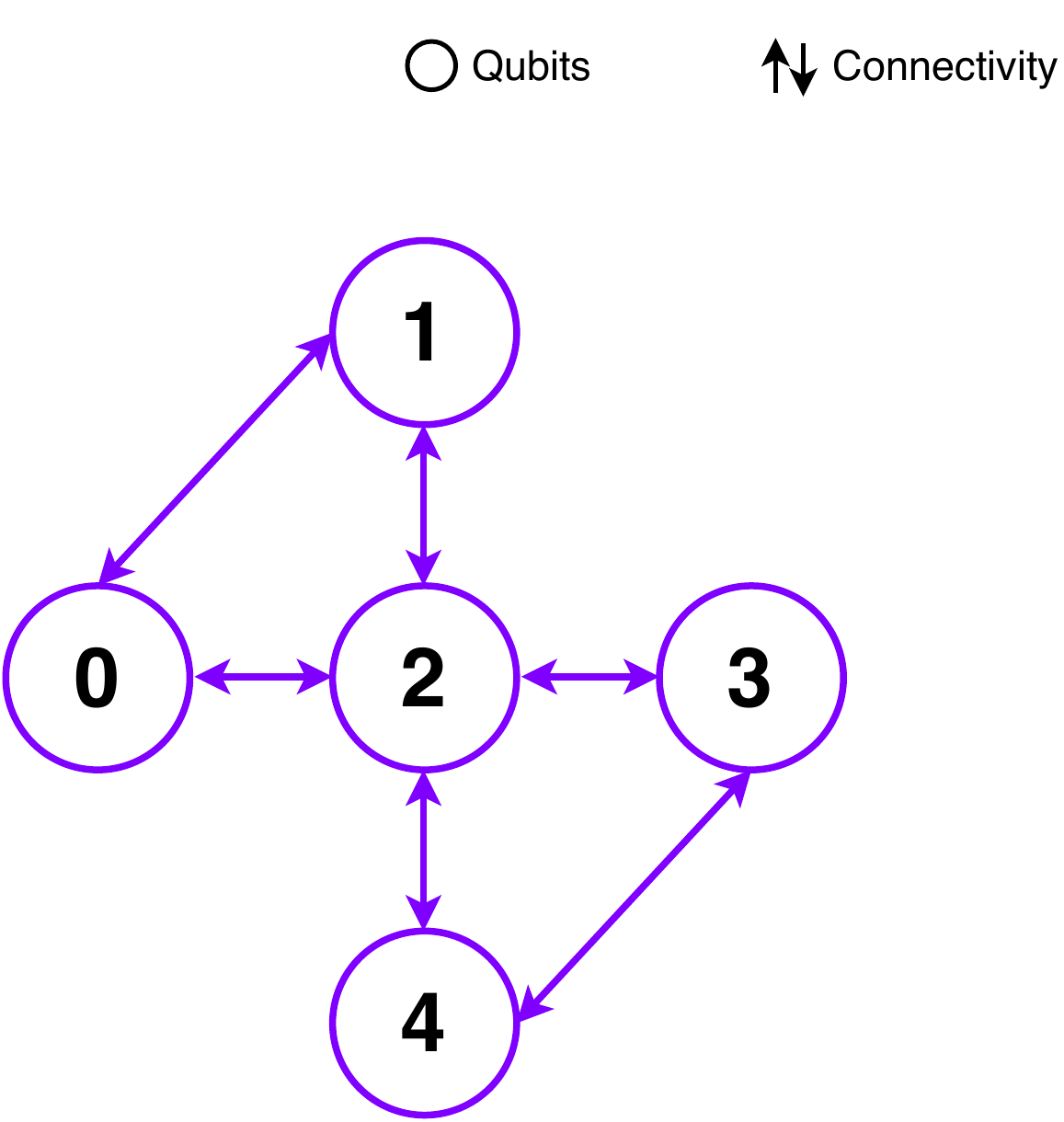}
  \caption{Hardware Architecture of ibmq\_5\_yorktown-ibmqx2.}
  \label{fig:ibmyorktown}
\end{figure}

This is also a real quantum hardware. It has got only 5 qubits. It supports the basics gates- u1, u2, u3, cx and id. Single-qubit U2 error rate is in between $4.802e-4$ and $5.927e-4$. CNOT error rate is in between $1.365e-2$ and $1.736e-2$. Figure \ref{fig:ibmyorktown} shows the architecture of the hardware. 
\subsection{ibmq\_armonk}
\begin{figure}[]
\centering
  \includegraphics[scale=0.5]{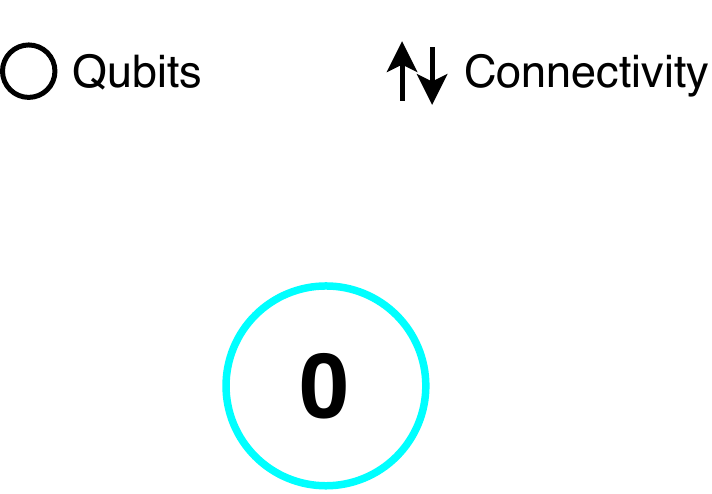}
  \caption{Hardware Architecture of ibmq\_armonk.}
  \label{fig:ibmarmonk}
\end{figure}

This is a real quantum hardware which has a single qubit. The design of this machine is shown in figure \ref{fig:ibmarmonk}. It supports the basics gates- id, u1, u2 and u3. Single-qubit U2 error rate is in between $6.985e-4$ and $6.985e-4$. CNOT error rate is in between $1.000e+0$ and $-1.000e+0$.  

\subsection{ibmq\_qasm\_simulator}
This device is a quantum simulator. It can simulate $32$ qubits. It supports the basic gates- u1, u2, u3, cx, cz, id, x, y, z, h, s, sdg, t, tdg, ccx, swap, unitary, initialize and kraus. 

%% file: tex/experiment.tex
\label{experiment}
In this section, the procedure of our experiments is described.  For this contribution we used IBM's quantum composer, publicly accessible at \url{https://quantum-computing.ibm.com/composer} to do some experiments. This composer gives us the opportunity to write programs on quantum machines. \\
To run those experiments we need to use Qiskit \cite{wiki2019qiskit} which is an open-source framework. It is basically used for quantum computing. It gives apparatuses for making and controlling quantum projects and running them on model quantum gadgets and test systems. It pursues the circuit model for widespread quantum calculation and can be utilized for any quantum equipment that pursues this model.\\
We have used Jupyter Notebook as our IDE. Anyone can use the online editor that IBM provides. We have also run our codes in the Google's colaboratory for some cases.

\subsection{The Harrow-Hassidim-Lloyd Algorithm}
Our goal of this contribution is to find out how the HHL algorithm performs on various matrices which we have found is not done earlier by anyone. To conduct this experiment, we used the platform, Google's Colaboratory and  Python to be our programming language. There are a few libraries available on Python programming language which help conduct quantum experiments. \href{https://qiskit.org/}{Qiskit}  is a  python library which is used for  IBM's Quantum Computer backend and available online. All of our codes for this experiment will be found at \href{https://github.com/ProtikNag/Experiment-with-HHL-Algorithm/blob/master/HHL.ipynb}{this github repository}. The  Qiskit official repository \cite{qiskit2019} has been of great assistance to us. \\ 
We have planned to observe the characteristics of the HHL algorithm from two different perspectives. One perspective is to show how the size of matrices affects the performance of the HHL algorithm. To do so, we have taken three matrices of $2 \times 2, 4 \times 4$ and $8 \times 8$ dimensions. Both diagonal and non-diagonal matrices have been experimented. We have also considered the sparsity of a matrix as a parameter to measure this algorithm. We aimed to understand how HHL works when the sparsity of a matrix increases or decreases. \\
It is  ensured that the matrix $A$ is a hermitian matrix.  Later, we have checked if the matrix is diagonal or not. Our assumption was  this algorithm can deal better with the system with diagonal matrices  than with non-diagonal matrices. We have tested different matrices with different sparsity. Again we have taken three $2\times2, 4\times4$ and $8\times8$ matrices with sparsity $0.5$ and $1$. The readers should be introduced to some particular metrics which are crucial to understand this experiment. 
\subsubsection{Probability}
Due to the probabilistic nature of the quantum computers, they cannot provide with the exact results. They run the same experiments for some number of times and show the probability of acquiring a particular result. This means they produce the probability of a result being true. We phrase the number of experiments as ``shots" and  the term ``probability"  refers to the highest probability of a result to be true.
\subsubsection{Fidelity}
``Fidelity" was first induced by Richard Jozsa \cite{jozsa1994fidelity}. Fidelity illustrates  how much we can rely on the result. Using Qiskit we can achieve the result of the classical computer  as well as the quantum computer simulation. Comparing the quantum computer's simulation with classical ones can produce the fidelity value. This value depicts the perfection of the proceeding. The more accurate result the quantum simulation provides, the closer the fidelity value becomes to $1$. Otherwise, the fidelity value declines to $0$.   We can represent  the results as quantum states and measure fidelity of those quantum states. State fidelity is the measure of how close two quantum states are to each other. If we simplify this, we can say that fidelity is the distance between two quantum states. It can be stated as the cosine of the smallest angle between two quantum states and also established as cosine similarity. This article \cite{fidelity} proves that  fidelity can be used as a standard parameter to distinguish two quantum states or to measure the similarity between two states. This metric has turned into one of the most universally used quantities to quantify the degree of similarity between quantum states. So we  regard fidelity to be a decisive parameter in our experiments too. 
\subsubsection{Sparsity}\
The  ``sparsity" of a matrix refers to the density of the generated matrix. Sparsity equal to $1$ means a full matrix with no element equals to $0$. Sparsity $0$ means a matrix with no non-zero items i.e.  each entry of the matrix will be zero. We assumed that the lesser the sparsity of the matrix is, the better the HHL performs. That means  with the lower sparsity matrix HHL performs better. The later section presents the comparisons necessary for this argument. 
\subsubsection{Circuit}
\begin{figure}[H]
\centering
  \includegraphics[scale=0.5]{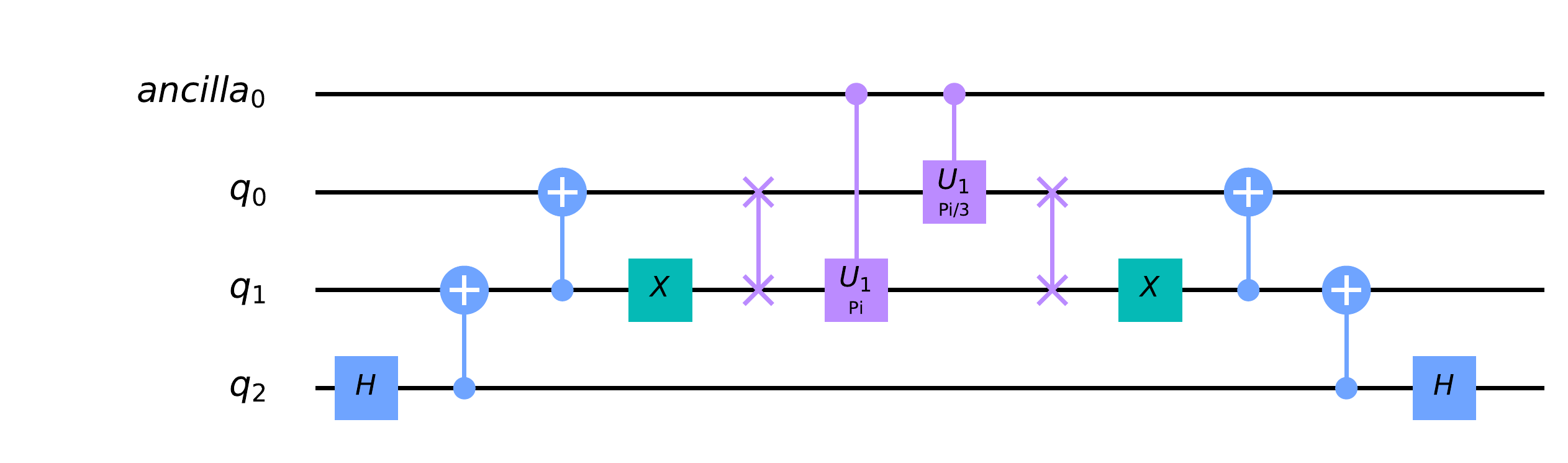}
  \caption{Circuit for Solving HHL Algortihm on a $2*2$ Matrix.}
  \label{fig:hhl}
\end{figure}

Different circuits combining different types of gates are necessary for solving distinct problems using quantum machines.  The circuit information consists of information about the depth  and the width of the circuit. The width of  a circuit also means the number of qubits used in a circuit. Circuit depth refers to the longest path required from the input to the output. Having a smaller quantum circuit requires less computational power. Figure \ref{fig:hhl} depicts the circuit description necessary for solving a $2*2$ matrix using the HHL algorithm.  We will show a comparative study based on the depth and width of the related quantum circuits in the next chapter. 
\subsubsection{Random Sparse Hermitian Matrix}
A matrix added to  its hermitian  is also a hermitian matrix\cite{yanofsky2008quantum}. So ascertaining our final matrix $B$ is hermitian, where $A$ is an arbitrarily generated sparse matrix, we used this equation
\begin{equation}
    B = A + A^\dag
\end{equation}
This makes the final matrix $B$ a hermitian matrix. We generated the random sparse matrix by using the SciPy library of Python.


\subsection{Quantum Support Vector Machine}

\subsubsection{Dataset}
In this part of our experiment, we have taken two beginner's datasets used for machine learning purposes. These are Iris dataset and the Breast Cancer dataset those can be easily accessed by the machine learning package Scikit-learn and are publicly available to use from \url{https://scikit-learn.org/stable/datasets/index.html}. 

\begin{figure}[]
\centering
  \includegraphics[scale=0.8]{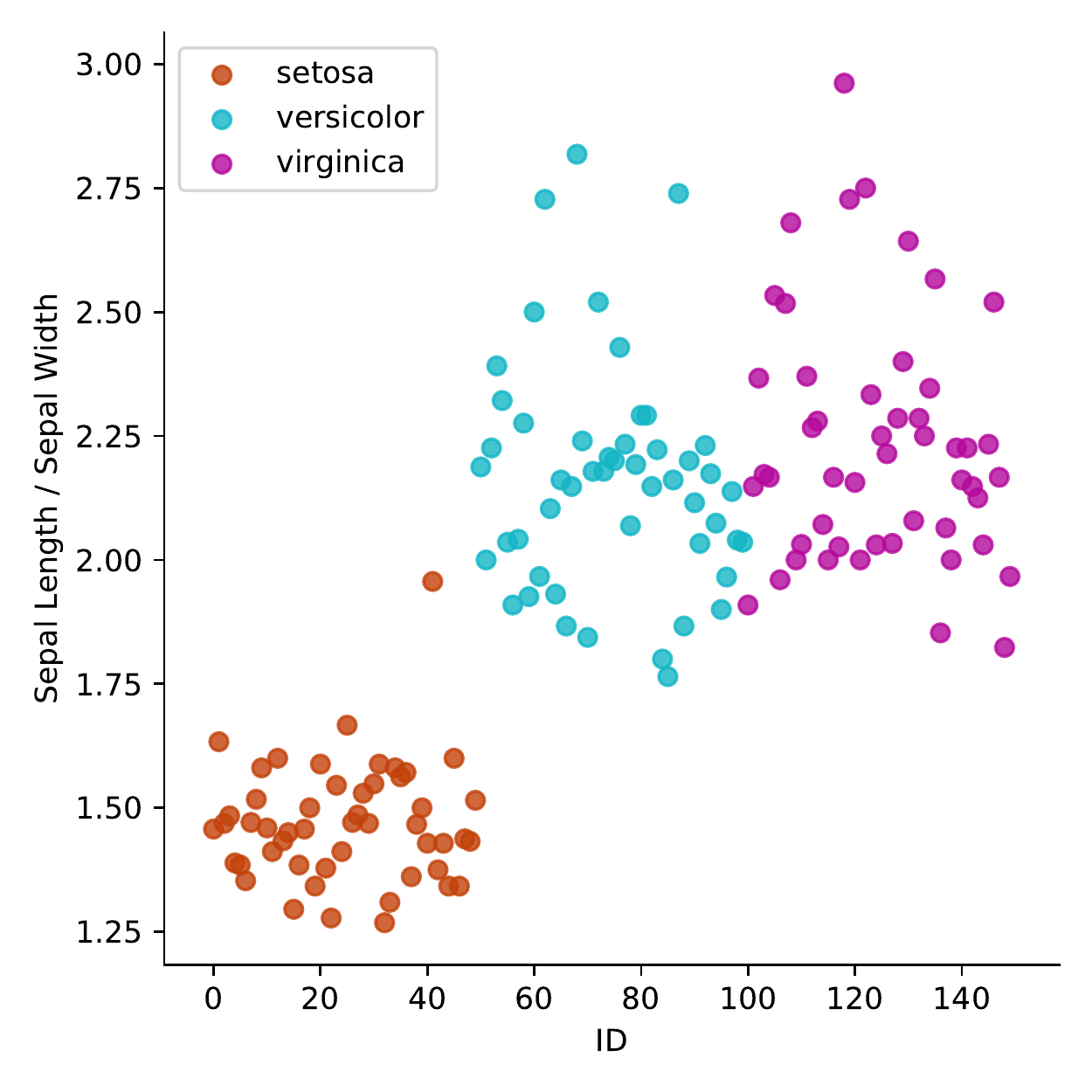}
  \caption{Visualization of the features of the Iris dataset.}
  \label{fig:iris}
\end{figure}

\paragraph{Breast Cancer Dataset}
The Breast Cancer dataset\cite{wolberg1992breast} is for the binary classification system. The target classes of this dataset are- Malignant and Benign. There are $569$ data-points in this dataset and each of them have $30$ features. We have separated the dataset into train and test set. The train set is kept small as we could not use so many data-points on the quantum support vector machine algorithm. We used $500$ data-points for test set and only 69 points for training.
\begin{figure}[]
\centering
  \includegraphics[scale=0.8]{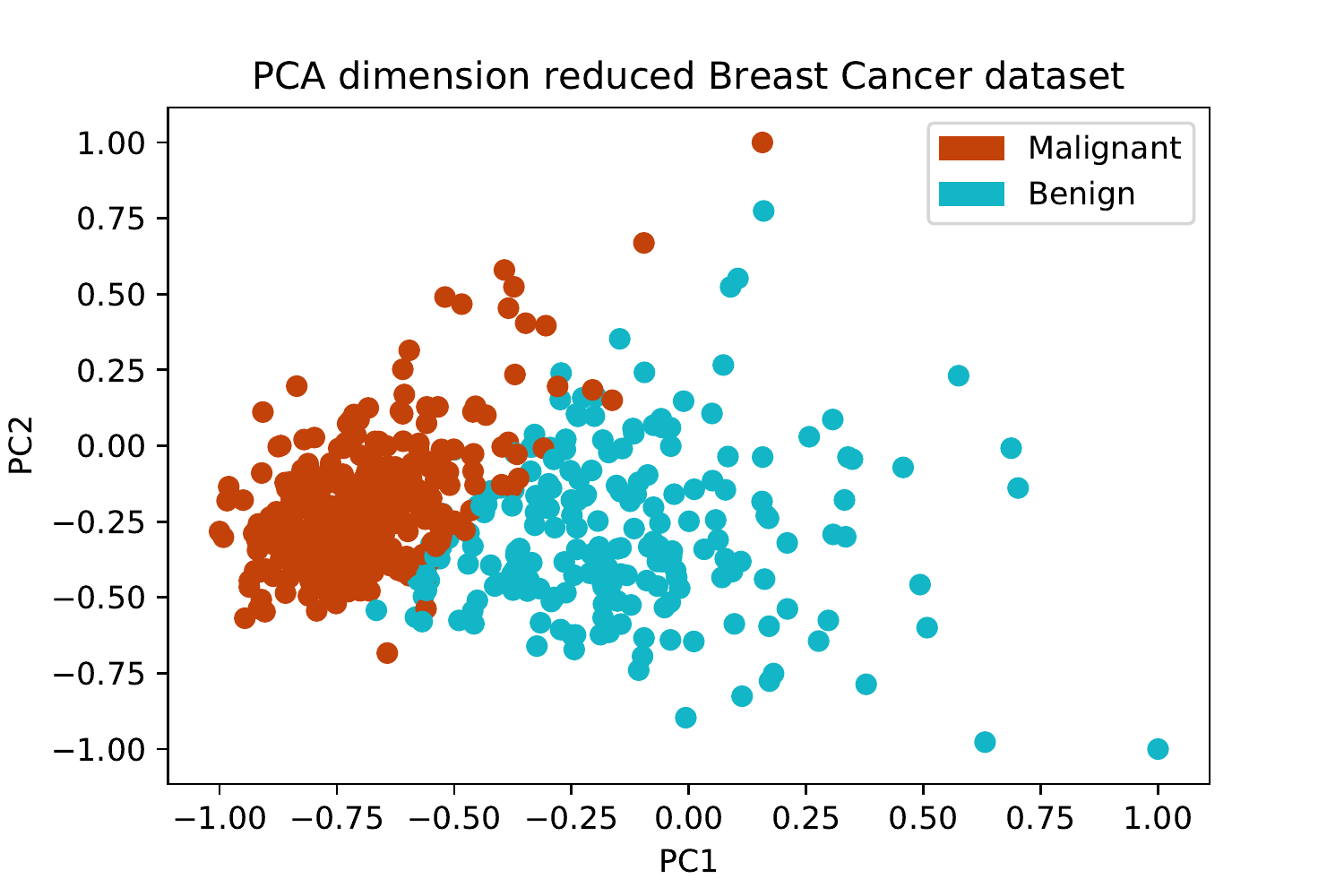}
  \caption{Visualization of the Breast Cancer dataset.}
  \label{fig:bc}
\end{figure}
\paragraph{Iris Dataset}
The  Iris dataset \cite{fisher1936uci} is for a multivariate classification problem. Iris dataset contains only $150$ data points and each of them has $4$ features. We have used 80\% of our data as the training data. Figure \ref{fig:iris} depicts the features of the Iris dataset.
\subsubsection{Preprocessing}
The datasets are preprocessed as they contain too many features and are not perfectly adjusted. For both the dataset we have applied Principal Component Analysis to diminish the dimensionality of the datasets. As more dimensionality induces more processing power, we have taken the decision to stand with the smaller number of features. For both datasets, we have lowered the dimensionality to $2$. Again, we have standardized the dataset around $0$ with unit variance. We have then scaled all features in between $-1$ to $+1$ so that the computation becomes simpler. 

\subsection{Implementation}
\subsubsection{Support Vector Machine} 
For both datasets, Support Vector Machine has been used as the operating engine for the classification problem. Support Vector Machine is used as a linear classifier that divides the hyperplane with maximum possible distance. We have selected linear kernel for SVM and $C=1$. 
\subsubsection{Quantum Support Vector Machine}
The Quantum part of this experiment on both datasets has been conducted in the QASM Simulator with $32$ qubits and depth $=2$. The shot numbers are $1, 2, 4, 8, 16, 32, 64, 128, 256, 512, 1024$. For feature mapping, second order expansion provided by Qiskit Aqua library is used. 

\subsection{Performance Evaluation}
For performance evaluation we have used four different metrics or parameters. They are accuracy, sensitivity, specificity and F1 score. The definitions of these metrics are given in the following:
\begin{align} 
Accuracy &= \frac{TP+TN}{TP+TN+FP+FN}\\
Sensitivity &= \frac{TP}{TP+FN}\\
Specificity &= \frac{TN}{TN+FP}\\
F1-score &= \frac{2TP}{2TP+FP+FN}
\end{align}
where TP, TN, FP, FN refers to true positive, true negative, false positive and false negative respectively. These values come from a confusion matrix described in table \ref{tab:confusion matrix}.
\begin{table}[]
\centering
\caption{Confusion Matrix for Binary Classification Problem.}
\label{tab:confusion matrix}
\begin{tabular}{cccc}
\multicolumn{2}{c}{\multirow{2}{*}{}}                 & \multicolumn{2}{c}{Predictions}                                          \\ \cline{3-4} 
\multicolumn{2}{c}{}                                  & \multicolumn{1}{|c|}{Positive}       & \multicolumn{1}{c|}{Negative}      \\ \cline{2-4} 
\multicolumn{1}{c|}{\multirow{2}{*}{\begin{tabular}[c]{@{}c@{}}Real\\ Labels\end{tabular}}} &
  \multicolumn{1}{c|}{Positive} &
  \multicolumn{1}{c|}{True Positive} &
  \multicolumn{1}{c|}{False Negative} \\ \cline{2-4} 
\multicolumn{1}{c|}{} & \multicolumn{1}{c|}{Negative} & \multicolumn{1}{c|}{False Positive} & \multicolumn{1}{c|}{True Negative} \\ \cline{2-4} 
\end{tabular}
\end{table}

%% file: tex/result.tex
\label{result}
\subsection{The  Harrow-Hassidim-Lloyd Algorithm }
\begin{figure}[]
\centering
  \includegraphics[scale=0.8]{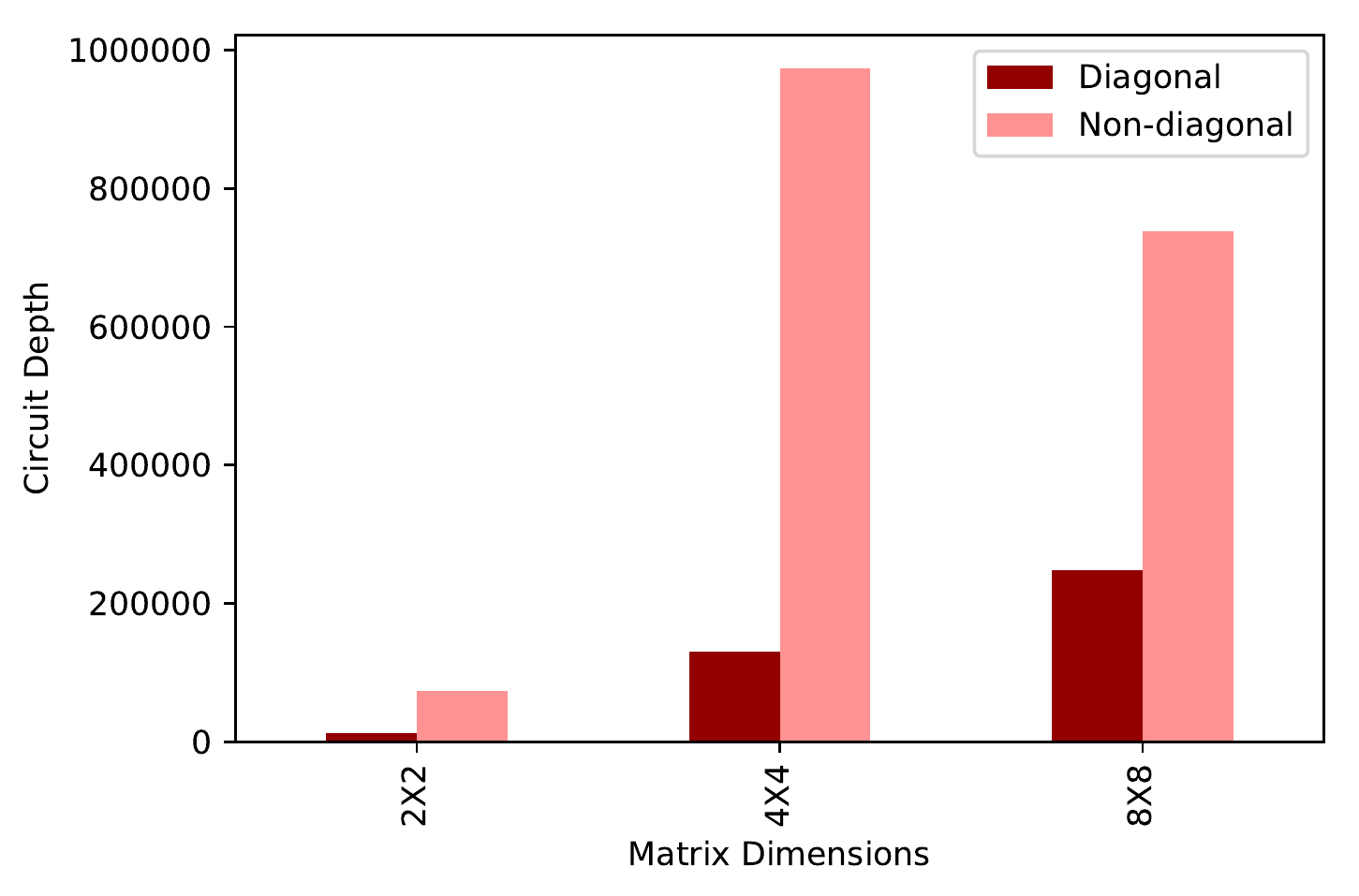}
  \caption{Quantum gate depth comparison of diagonal and non-diagonal matrices.}
  \label{fig:Depth}
\end{figure}

This section illustrates the results of the conducted experiments. As the preceding chapter articulates, our approach is to figure out the algorithm based on two distinct norms which we introduced  earlier. \\
Our initial point of concern has been  whether a matrix is diagonal or not. We have assumed that a diagonal matrix may be simpler to analyze than a non-diagonal matrix. Statistics that we have found from our experiments reinforce our hypothesis. After experimenting we have discovered that the fidelity of  matrices of various dimensions does not vary on the matrix being diagonal or not. This  means it can solve the system of the linear equation almost flawlessly. But the actual difference comes to the sight when we become concern about the probability and the circuit information. Probability dramatically falls if a matrix is non diagonal. 
\begin{figure}[]
\centering
  \includegraphics[scale=0.8]{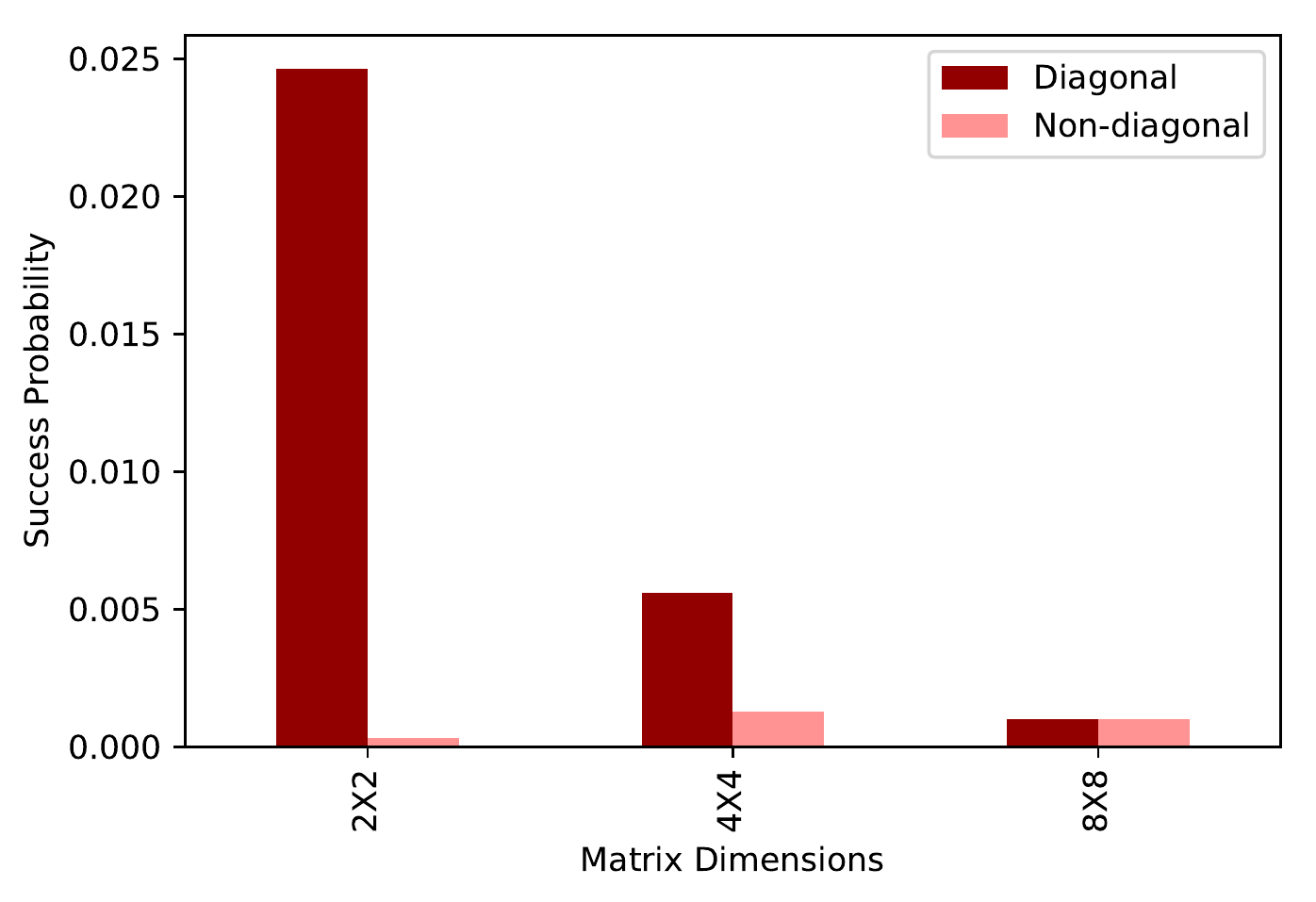}
  \caption{Success probabilities of diagonal and non-diagonal matrices.}
  \label{fig:Probability}
\end{figure}
Figure \ref{fig:Probability}  explains the probability comparison. The dropped probability depicts that  additional unnecessary solutions have been generated by the quantum machine and therefore the  probability of the actual solution is declined. So, it leads to the wastage of computational power. \\
The circuit information also shows the evidence of misuse of computing power. As the size of the matrix enhances, the size of the circuit increases as well. We somewhat expect this increment. But surprisingly, the circuit size also expands when the HHL algorithm deals with the non-diagonal matrices.\\
\begin{figure}[]
    \centering
  \includegraphics[scale=0.8]{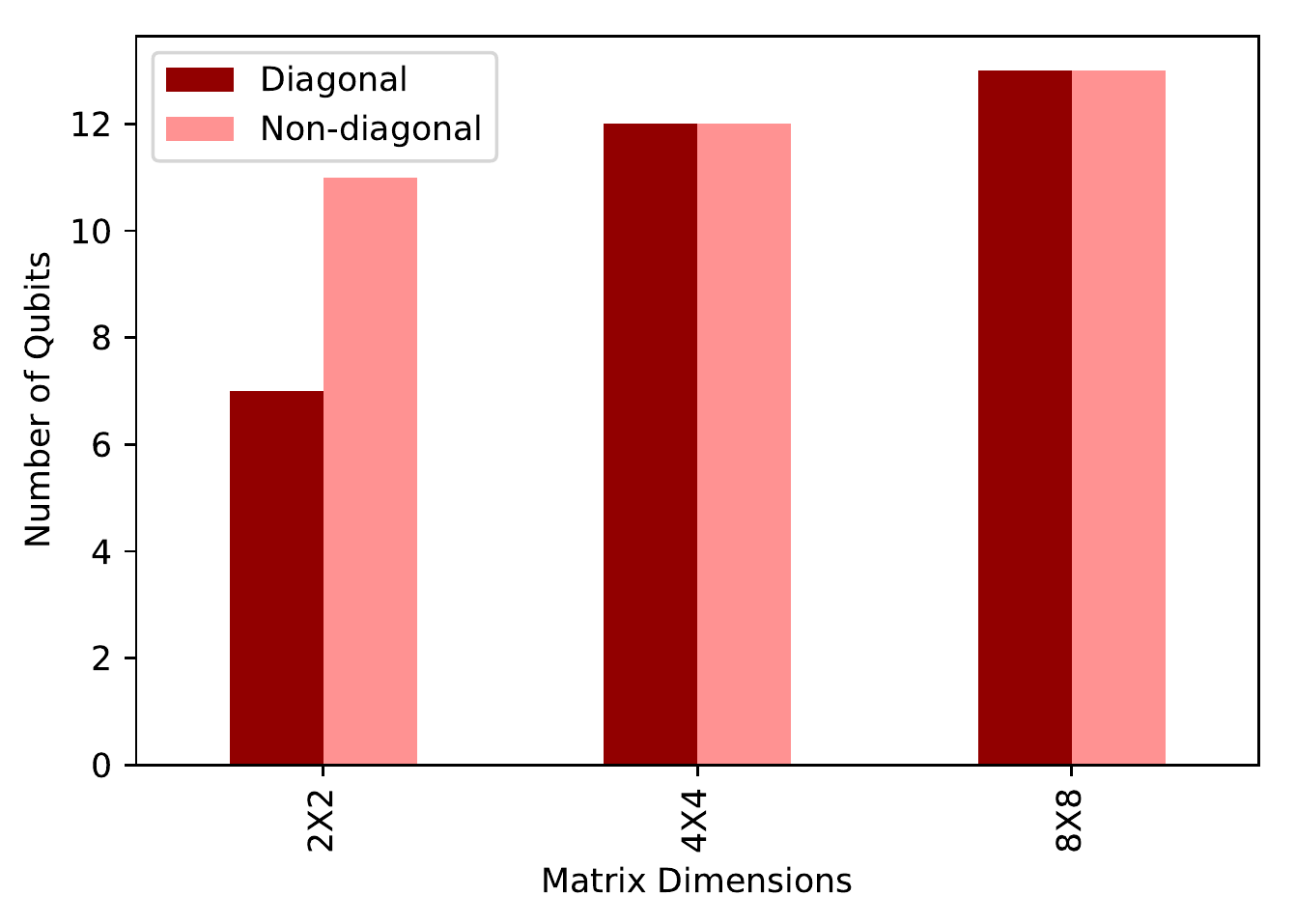}
  \caption{Number of qubits needed for diagonal and non-diagonal matrices.}
  \label{fig:Width}
\end{figure}
Figure \ref{fig:Width}  exhibits the circuit's width which means the number of qubits both for diagonal and non-diagonal matrices. For $4\times4$ and $8\times8$ matrices the width of the circuit does not increase. But when we plot the depth of the circuit, the contrast becomes clear. Figure \ref{fig:Depth} shows the difference. 

From this discussion, it assures our conjecture that the system of the linear equation with non-diagonal matrices take much higher computational power and considerable effort to be solved. Quantum machines require more circuit width and depth to execute the computation. Also, it provokes some invalid outputs to diminish the  probability of actual output even though the fidelity remains approximately constant. Table \ref{tab: table1} shows the accurate comparison between diagonal and non-diagonal matrices. 

\begin{table}[]
\centering
\caption{ Overall Comparison between Diagonal and Non-diagonal Matrices.}
\label{tab: table1}
\begin{tabular}{|c|c|c|c|}
\hline
\textbf{\begin{tabular}[c]{@{}c@{}}Dimension\\ of the \\ Matrix\end{tabular}} &
  \textbf{Parameters} &
  \textbf{Diagonal Matrix} &
  \textbf{\begin{tabular}[c]{@{}c@{}}Non-diagonal\\ Matrix\end{tabular}} \\ \hline
\multirow{4}{*}{2 x 2} & Probability                                              & 0.024630 & 0.000316 \\ \cline{2-4} 
                       & Fidelity                                                 & 0.999389 & 0.999996 \\ \cline{2-4} 
                       & \begin{tabular}[c]{@{}c@{}}Circuit \\ Width\end{tabular} & 7        & 11       \\ \cline{2-4} 
                       & \begin{tabular}[c]{@{}c@{}}Circuit\\ Depth\end{tabular}  & 12256    & 73313    \\ \hline
\multirow{4}{*}{4 x 4} & Probability                                              & 0.005583 & 0.001285 \\ \cline{2-4} 
                       & Fidelity                                                 & 1.0      & 0.999994 \\ \cline{2-4} 
                       & \begin{tabular}[c]{@{}c@{}}Circuit \\ Width\end{tabular} & 12       & 12       \\ \cline{2-4} 
                       & \begin{tabular}[c]{@{}c@{}}Circuit \\ Depth\end{tabular} & 130723   & 973521   \\ \hline
\multirow{4}{*}{8 x 8} & Probability                                              & 0.000982 & 0.001002 \\ \cline{2-4} 
                       & Fidelity                                                 & 0.999992 & 0.999997 \\ \cline{2-4} 
                       & \begin{tabular}[c]{@{}c@{}}Circuit \\ Width\end{tabular} & 13       & 13       \\ \cline{2-4} 
                       & \begin{tabular}[c]{@{}c@{}}Circuit\\  Depth\end{tabular} & 248342   & 738340   \\ \hline
\end{tabular}
\end{table}

\begin{figure}[]
\centering
  \includegraphics[scale=0.8]{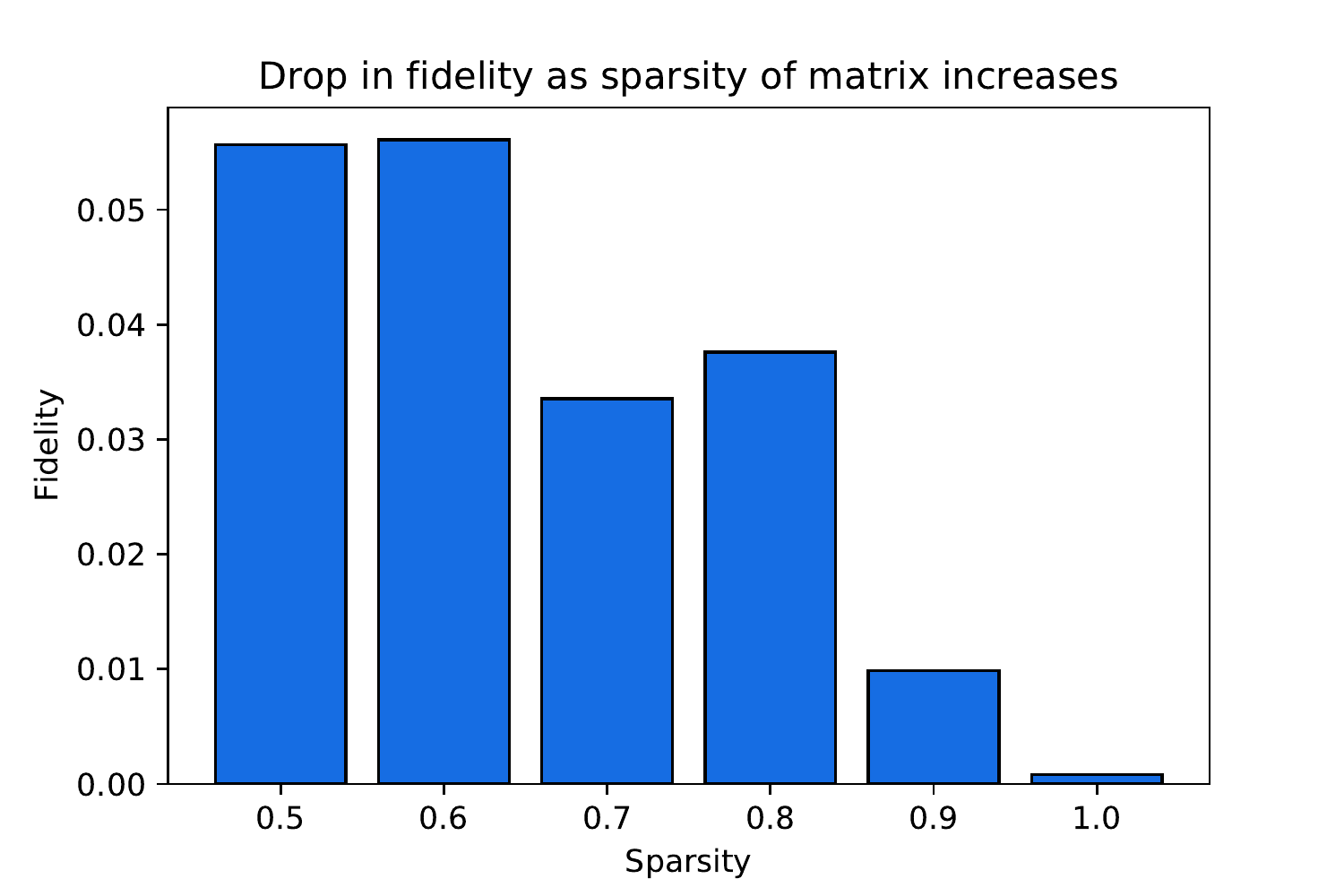}
  \caption{Drop in Fidelity with respect to Sparsity. }
  \label{fig:drop fidelity}
\end{figure}

Up to this point, the result is satisfactory in spite of taking extended computing capability because it does not influence the fidelity. The output is approximately identical as the classical output. The table \ref{tab: table1} demonstrates that the fidelity is relatively uninterrupted which is over $99\%$ in every instance. But the fidelity dramatically drops when the experiment is based on arbitrary sparse hermitian matrix. When we observe  the dimensions of the matrices, we can see that the fidelity shrinks as the dimensions increase. We can solve the matrix of size $2\times2$ with fidelity $0.422602$. But it falls down to $0.000011$ when the size becomes $4\times4$. This fidelity  tends to zero which implies we cannot rely on this outcome and so is of no benefit.\\
\begin{figure}[]
\centering
  \includegraphics[scale=0.8]{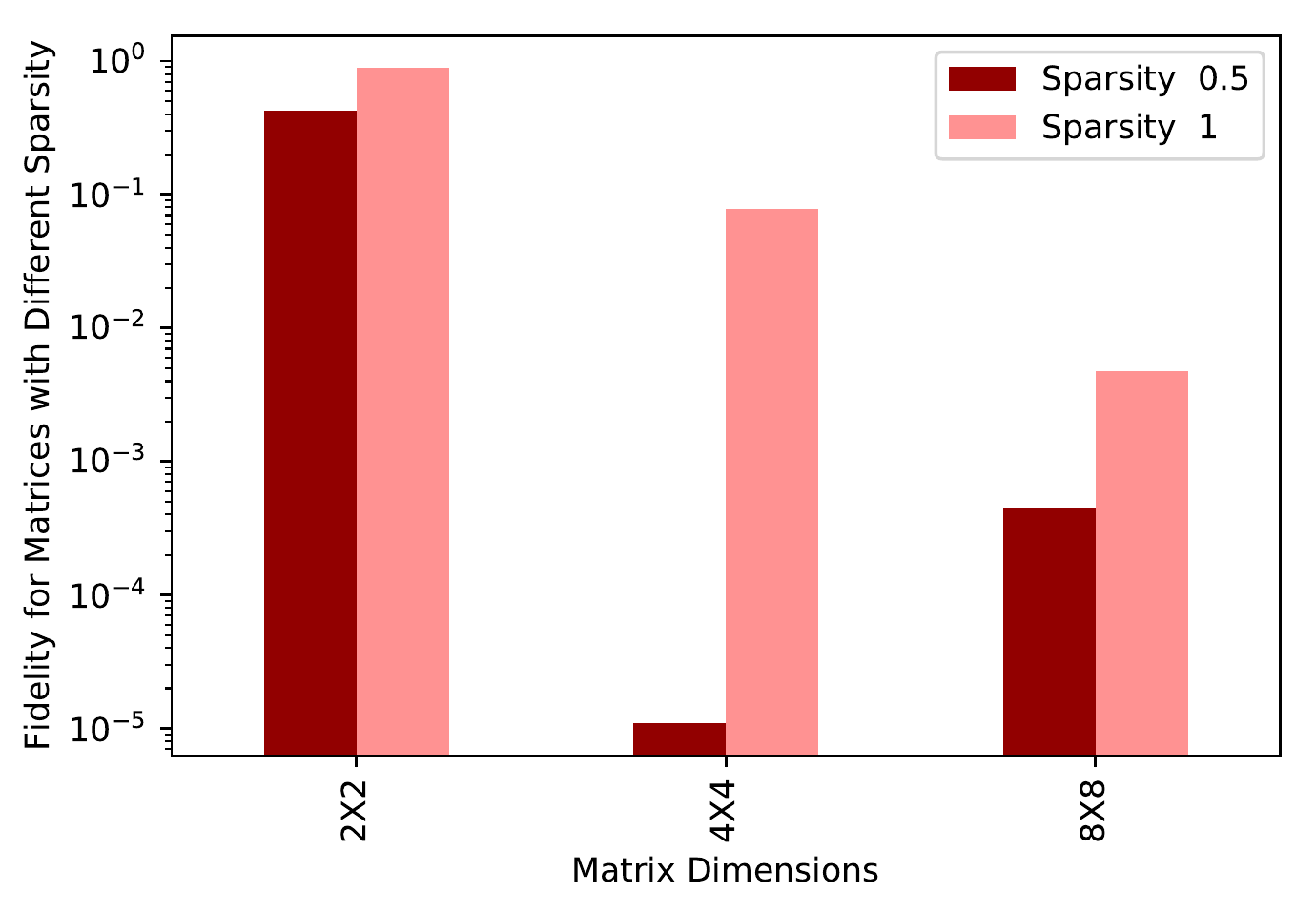}
  \caption{Diminishing Fidelity with Random Sparse Hermitian Matrix.}
  \label{fig:Fid}
\end{figure}
The figure \ref{fig:Fid} illustrates how the fidelity is damaged when we want to solve the system of the linear equation with random sparse hermitian matrix with various densities. This figure shows the fidelity value in logarithmic scale. \\
From table \ref{tab:table2} we see that the circuit size deepens as the matrix grows bigger. We  anticipated this to a small extent. But here, we can observe that the sparsity has no effect on the circuit size. With both of the sparsity $0.5$ and sparsity $1$, the circuit size remains the same.\\
\begin{table}[]
\centering
 \caption{ Overall Comparison between Sparsity $0.5$ and Sparsity $1$ matrices.}
\label{tab:table2}
\begin{tabular}{|c|c|c|c|}
\hline
\textbf{\begin{tabular}[c]{@{}c@{}}Dimension\\ of the \\ Matrix\end{tabular}} &
  \textbf{Parameters} &
  \textbf{\begin{tabular}[c]{@{}c@{}}Sparsity\\ = 0.5\end{tabular}} &
  \textbf{\begin{tabular}[c]{@{}c@{}}Sparsity \\ = 1\end{tabular}} \\ \hline
\multirow{4}{*}{2 x 2} & Probability                                              & 0.099122 & 0.028234 \\ \cline{2-4} 
                       & Fidelity                                                 & 0.422602 & 0.883097 \\ \cline{2-4} 
                       & \begin{tabular}[c]{@{}c@{}}Circuit \\ Width\end{tabular} & 7        & 7        \\ \cline{2-4} 
                       & \begin{tabular}[c]{@{}c@{}}Circuit\\ Depth\end{tabular}  & 30254    & 30254    \\ \hline
\multirow{4}{*}{4 x 4} & Probability                                              & 0.135214 & 0.146572 \\ \cline{2-4} 
                       & Fidelity                                                 & 0.000011 & 0.077870 \\ \cline{2-4} 
                       & \begin{tabular}[c]{@{}c@{}}Circuit \\ Width\end{tabular} & 8        & 8        \\ \cline{2-4} 
                       & \begin{tabular}[c]{@{}c@{}}Circuit \\ Depth\end{tabular} & 165262   & 165262   \\ \hline
\multirow{4}{*}{8 x 8} & Probability                                              & 0.087196 & 0.152713 \\ \cline{2-4} 
                       & Fidelity                                                 & 0.000453 & 0.004737 \\ \cline{2-4} 
                       & \begin{tabular}[c]{@{}c@{}}Circuit \\ Width\end{tabular} & 9        & 9        \\ \cline{2-4} 
                       & \begin{tabular}[c]{@{}c@{}}Circuit\\  Depth\end{tabular} & 801281   & 801281   \\ \hline
\end{tabular}
\end{table}
There is another feature the HHL algorithm shows. The fidelity of the solution diminishes as the sparsity of the matrix increases. To investigate this characteristic, we have chosen matrices of $0.5, 0.6, 0.7, 0.8, 0.9$ and $1.0$ sparsity. To make our result more acceptable, we have taken $3$ random sparse matrix for each of these sparsities and solved the system of linear equation. The figure \ref{fig:drop fidelity} demonstrates how fidelity decreases with density. This causes this algorithm more susceptible.\\
After all the experiments are done, it is time to sum up all the results. As this \cite{xuan2016cr04} contribution summarizes, HHL algorithm is for linear systems of equations when the input $n\times n$ matrix is $s$-sparse and well-conditioned. And most notably, the solutions are not exact as we only obtain an approximate solution for this problem. From our experiments we have identified that this HHL algorithm bears some further limitations too. One of the most remarkable disadvantage of this algorithm is it requires more computation capability when a matrix from a system of the linear equation is non-diagonal. \\
The most significant disadvantage of HHL algorithm is it cannot solve the random sparse hermitian matrix. What it produces as result for random sparse hermitian matrix is not accurate at all. Though we hypothesized that this algorithm may perform better with the matrices with lower sparsity, we have now come to find that it indeed does not work on random sparse hermitian matrix. It produces faulty outputs when the experiments are carried out with random sparse hermitian matrices. The positive side is that it does not consume extra computational power to solve the matrices with higher sparsity as we have found the circuit size has remained the same. \\
In spite of being a remarkably powerful algorithm in the field of quantum computing, the HHL algorithm possesses some limitations. In this contribution, we have investigated to find out some of these limitations. Our aim was to experience the behavior of the HHl algorithm when it encounters some unusual situations. This contribution presents the urgency of tuning the HHL algorithm up and design an updated or new algorithm to cope with these critical cases. 

\subsection{Quantum Support Vector Machine}
The performance of the quantum computer has not been up to the mark. The classical interpretation of the Support Vector Machine algorithm has outperformed the quantum one. But despite possessing plenty of obstacles, the quantum variant of the support vector machine has not performed bad. In this section,  a comparative investigation between the classical and quantum variant of the SVM algorithm on two distinctive datasets mentioned in the previous section is displayed. At first, the performance comparison is depicted when we have conducted the experiment on the quantum computer just for once. Afterwards the information about how the performance steps up with the quantity of experiments increases is shown. 
\subsubsection{Classical vs Quantum SVM}
\begin{figure}[]
\centering
  \includegraphics[scale=0.9]{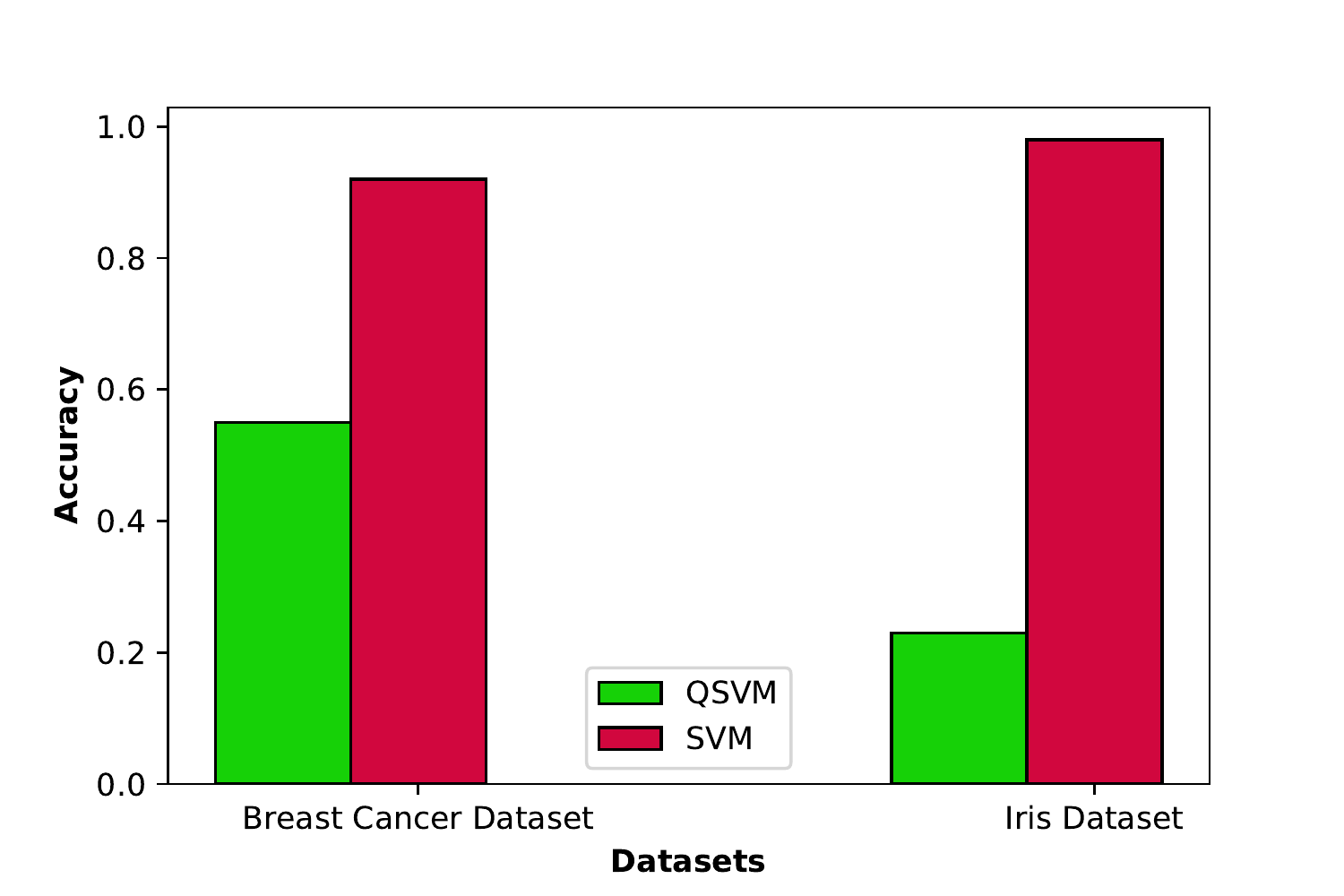}
  \caption{Accuracy Comparison for Classical and Quantum SVM for both datasets.}
  \label{fig:qsvm}
\end{figure}

On both of the datasets, the classical algorithm performed better. The figure \ref{fig:qsvm} illustrates the difference. For the Breast Cancer dataet, the Quantum SVM gives us only $55\%$ accuracy after we take simply one shot. The classical subpart, for this dataset, presents a remarkable $92\%$ of accuracy. Again, on the Iris Dataset, QSVM gives a worse output. It can achieve only $23\%$ of accuracy after one shot. Certainly, this is not a progress. It is far away from the state-of-the-art performance. This result is some sort of random predictions. The classical one can produce a stunning $98\%$ accuracy without any kind of kernel tricks. So, there are many opportunities for upgrading the performance. 

\subsubsection{QSVM with Shots Increased}
The number of shots is the total of conducted experiments on a quantum system. We have observed that conducting the experiments several times on the quantum machine provides us a stronger performance for both of the datasets. Though for a specific period it declined for both of the datasets, but overall performance is increased. The dilemma is that taking more shots can bring about more processing power, which is not anticipated. So, there is always a compensation between the efficiency and the number of shots.
\begin{figure}[]
\centering
  \includegraphics[scale=0.8]{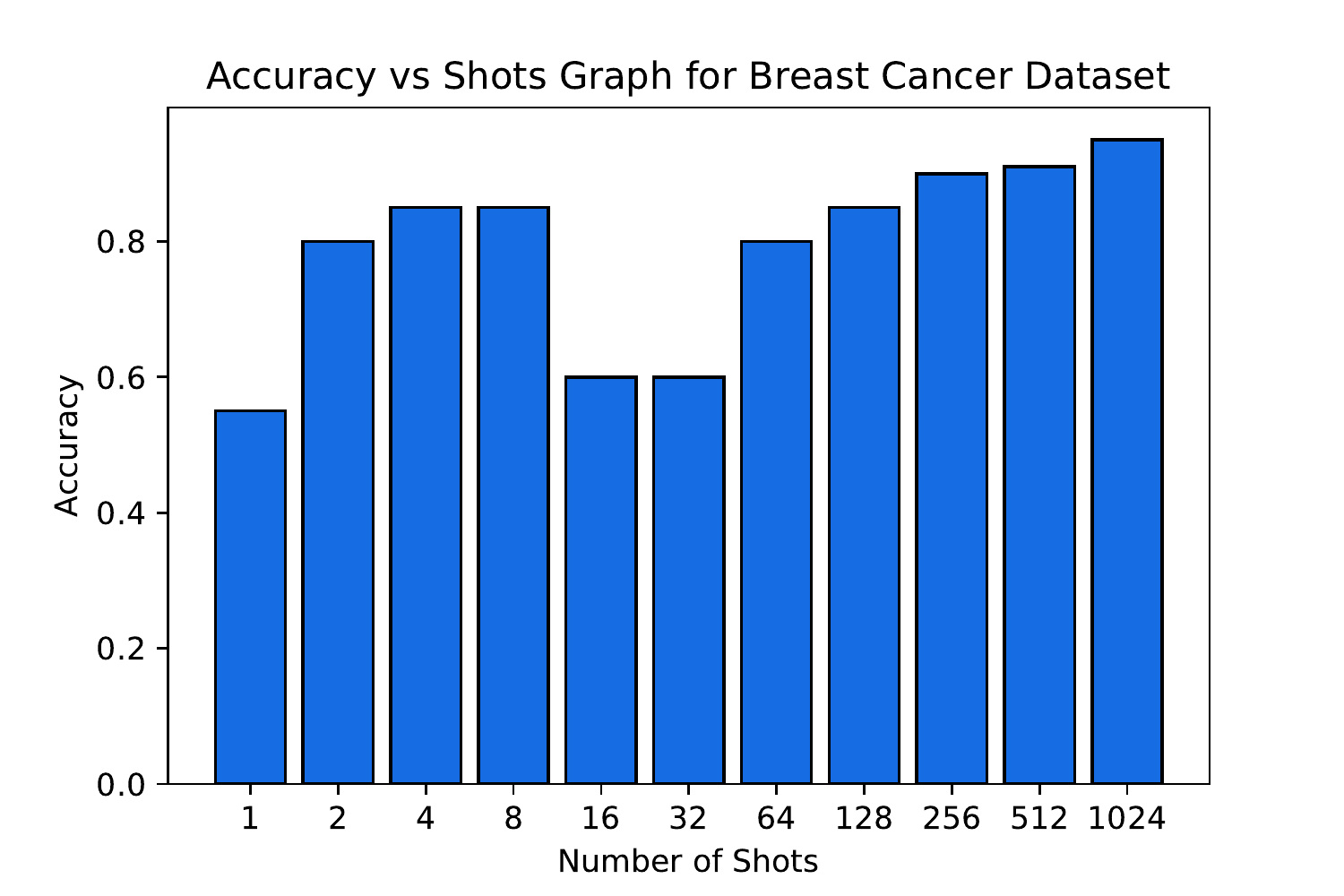}
  \caption{Accuracy vs shots on Breast Cancer Dataset.}
  \label{fig:bcd shots}
\end{figure}

\begin{figure}[]
\centering
  \includegraphics[scale=0.8]{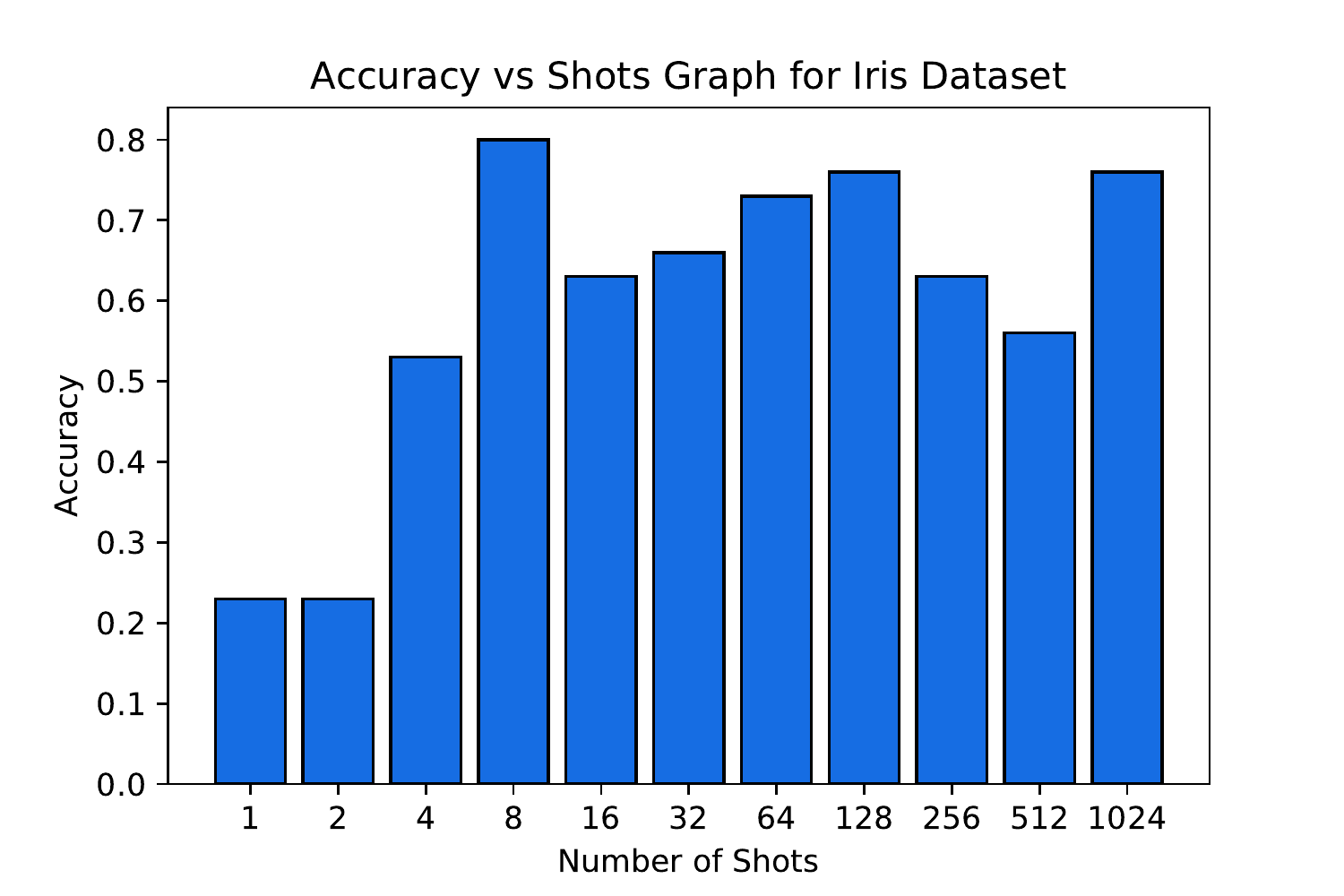}
  \caption{Accuracy vs shots on Iris Dataset.}
  \label{fig:iris shots}
\end{figure}
The results of the increased shots have been demonstrated in figure \ref{fig:bcd shots} and \ref{fig:iris shots}. Though this accuracy has some ups and downs, the performance has increased noticeably. From this experiment, we can come to a settlement that, QSVM shows better performance on binary classification problem than on multi-label classification problem. With sufficient quantity of shots, the performance can be as satisfying as classical computer for binary classification problem. But there are many fields left for multi-label classification where further consideration is demanded. 

\begin{figure}[H]
     \centering
     \begin{subfigure}[b]{0.6\linewidth}
         \centering
         \includegraphics[width=\linewidth]{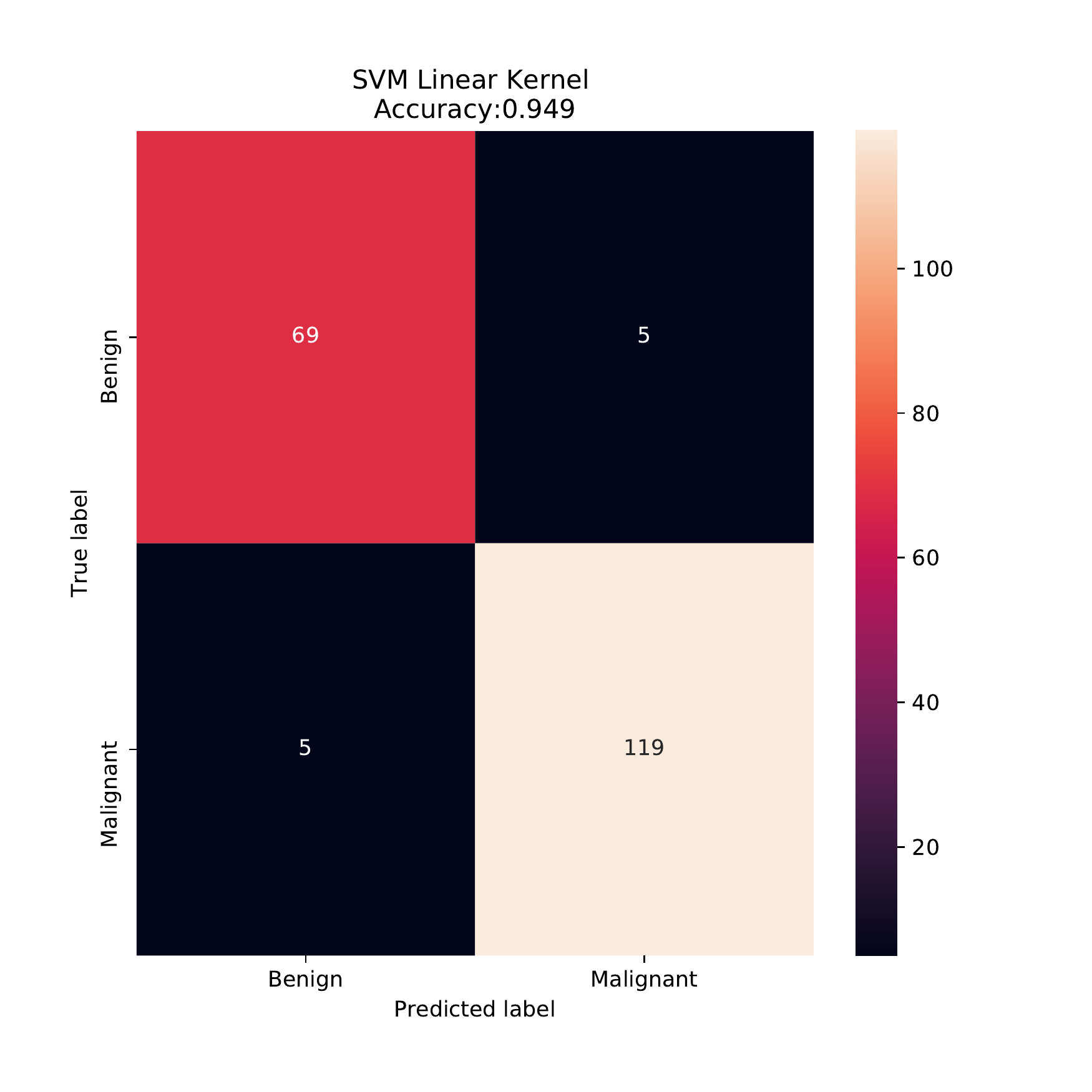}
         \caption{Confusion Matrix of SVM on Breast Cancer Dataset.}
         \label{fig:svm bcd matrix}
     \end{subfigure}
     \hfill
     \begin{subfigure}[b]{0.6\linewidth}
         \centering
         \includegraphics[width=\linewidth]{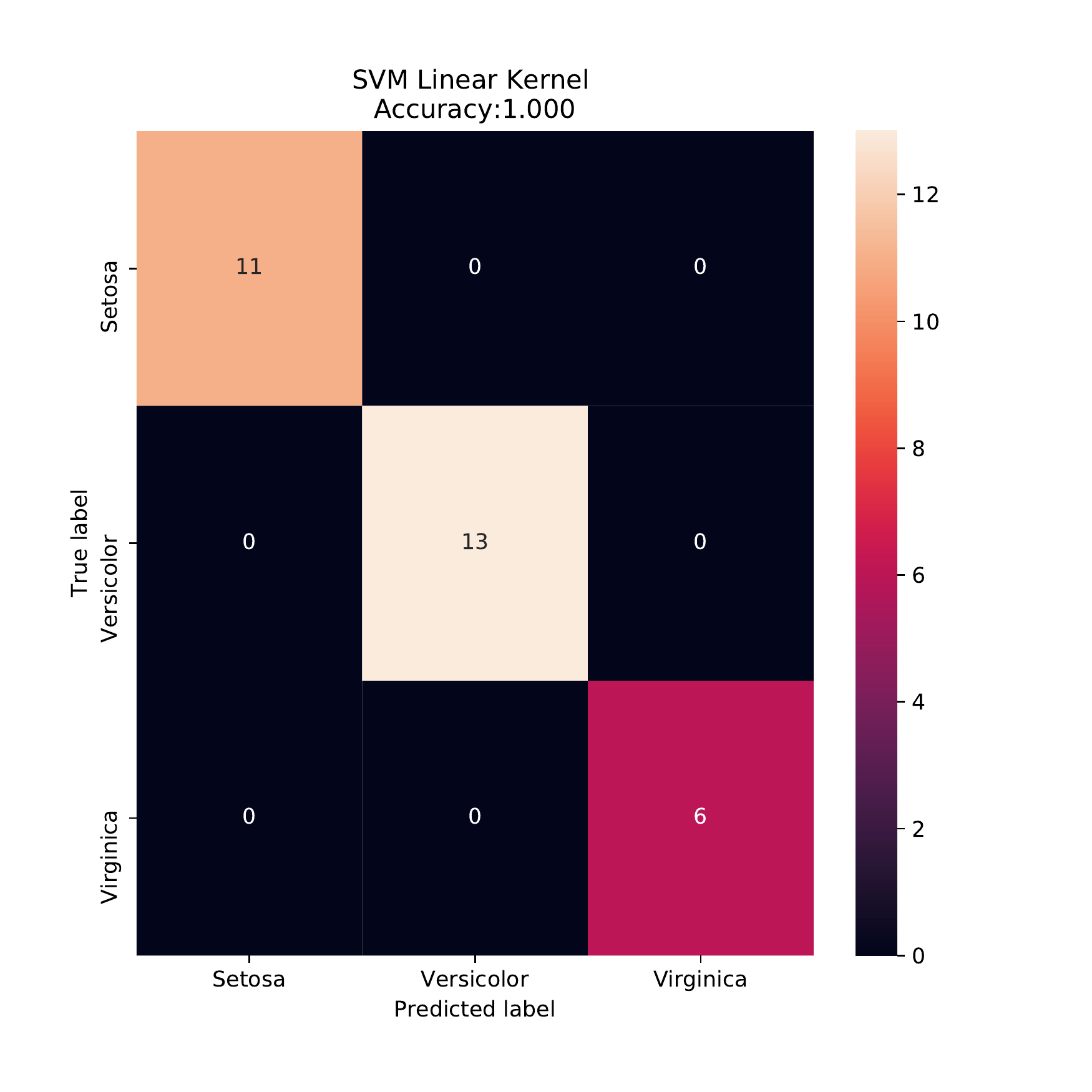}
         \caption{Confusion Matrix of SVM on Iris Dataset.}
         \label{fig:iris svm matrix}
     \end{subfigure}
        \caption{Confusion Matrices from Classical SVM experiments.}
        \label{fig:svm matrices}
\end{figure}

\begin{figure}[H]
     \centering
     \begin{subfigure}[b]{0.6\linewidth}
         \centering
         \includegraphics[width=\linewidth]{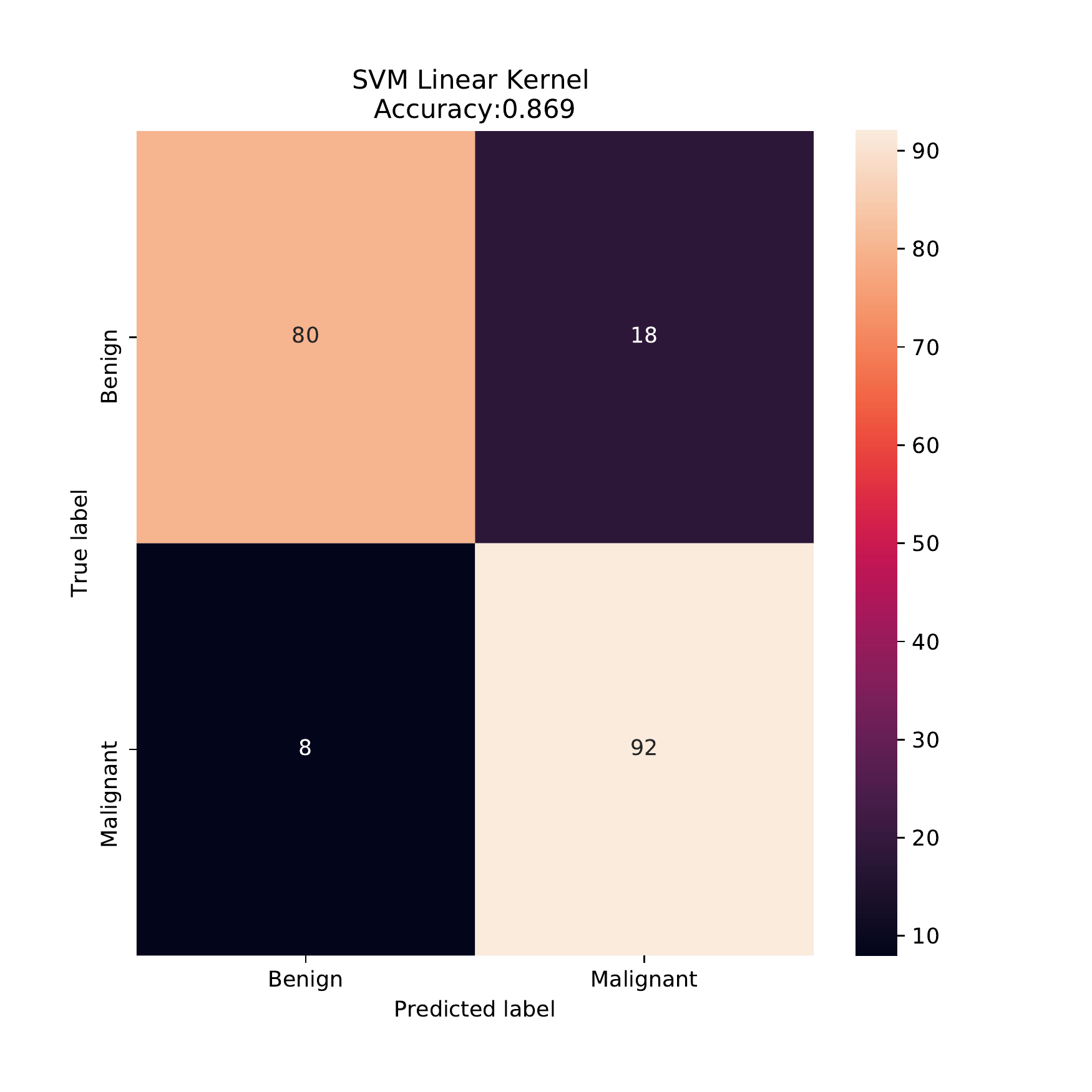}
         \caption{Confusion Matrix of QSVM on Breast Cancer Dataset.}
         \label{fig:qsvm bcd matrix}
     \end{subfigure}
     \hfill
     \begin{subfigure}[b]{0.6\linewidth}
         \centering
         \includegraphics[width=\linewidth]{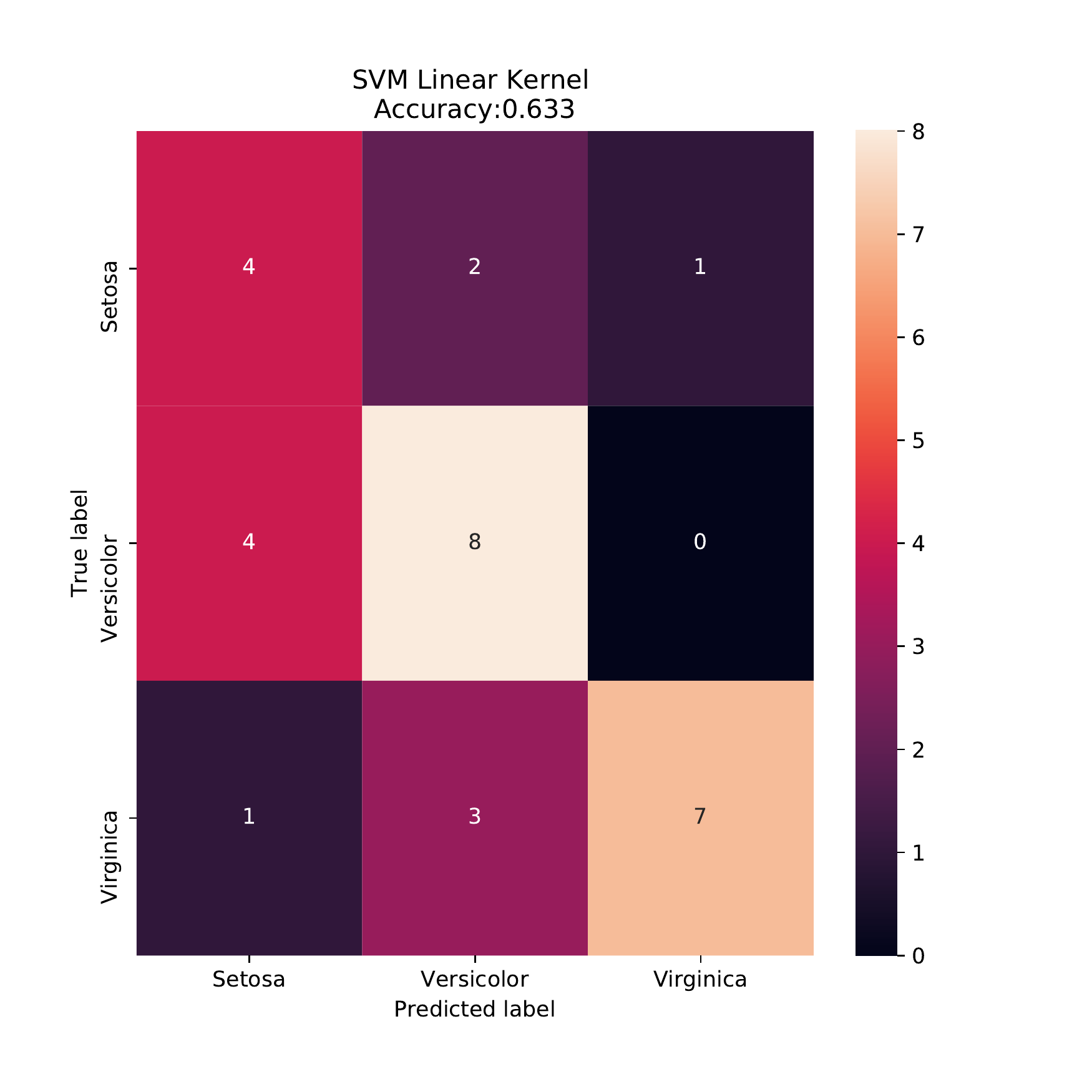}
         \caption{Confusion Matrix of QSVM on Iris Dataset.}
         \label{fig:iris qsvm matrix}
     \end{subfigure}
        \caption{Confusion Matrices from Quantum SVM experiments.}
        \label{fig:qsvm matrices}
\end{figure}

The confusion matrices shown in \ref{fig:svm matrices} and \ref{fig:qsvm matrices} give the idea of exact results that have been found through the experiments. To analyze our results we have generated 4 confusion matrices for both of the datasets. On both of them, we have applied SVM and QSVM algorithm. To obtain the finest result, the number of shots on which the QSVM performed best are chosen, which are $1024$ or the Breast Cancer dataset  and  $8$ for Iris dataset. As from the figure \ref{fig:bcd shots} and figure \ref{fig:iris shots} shows the best number of shots for these two datasets.

Let us go into a bit deep into the confusion matrix for the Breast Cancer dataset. We have taken 198 samples for evaluation purpose. SVM kernel has produced $94.9\%$ of accuracy. So, most of the cases have been recognized as True Positives and True Negatives. 69 samples have been designated as Benign class and 119 samples have been detected as Malignant class. For a total of 188 samples out of 198 has been perfectly classified. Remaining 10 specimens have been labeled incorrectly. On the other hand, the Quantum Support Vector Machine kernel produces $86.9\%$ of accuracy. Total 172 samples have been classified accurately. Among them 80 samples have been chosen as Benign class and remaining 92 have been classified as Malignant class. 

On the Iris dataset, the QSVM kernel has not performed that much satisfactory. This dataset includes only 150 samples. This can be an explanation of poor performance. We have taken 30 of them for testing purpose. The SVM kernel has performed adequately on this dataset. SVM can produce $100\%$ of accuracy on these 30 samples. So all of them as classified as True Positives and True Negatives. Again, the QSVM kernel has not performed that much good on this dataset. Total 19 samples out of 30 has been classified precisely. This demonstrates that there are many other opportunities of investigation in multi-label classification problem. 

Table \ref{tab:result} presents different features of the confusion matrices such as- Accuracy, F1 Score, Sensitivity, Specificity  and compares them.

\begin{table}[]
\centering
\caption{Comparison of Performance of Prediction between SVM and QSVM on Two Datasets.}
\label{tab:result}
\begin{tabular}{ccccccc}
\hline
\textbf{Dataset} &
  \textbf{Classifier} &
  \textbf{\begin{tabular}[c]{@{}c@{}}Name of \\ classes\end{tabular}} &
  \textbf{Accuracy} &
  \textbf{Sensitivity} &
  \textbf{Specificity} &
  \textbf{F1 Score} \\ \hline
\multicolumn{1}{c|}{\multirow{2}{*}{Breast Cancer}} &
  \multicolumn{1}{c|}{SVM} &
  \multicolumn{1}{c|}{\multirow{2}{*}{\begin{tabular}[c]{@{}c@{}}Benign \\ and  Malignant\end{tabular}}} &
  0.9495 &
  0.9324 &
  0.9597 &
  0.9324 \\ \cline{2-2}
\multicolumn{1}{c|}{} & \multicolumn{1}{c|}{QSVM} & \multicolumn{1}{c|}{}           & 0.8687 & 0.8162 & 0.9224 & 0.8623 \\ \hline
\multicolumn{1}{c|}{\multirow{6}{*}{Iris}} &
  \multicolumn{1}{c|}{\multirow{3}{*}{SVM}} &
  \multicolumn{1}{c|}{Setosa} &
  \multirow{3}{*}{1.000} &
  1.000 &
  1.000 &
  1.000 \\ \cline{3-3}
\multicolumn{1}{c|}{} & \multicolumn{1}{c|}{}     & \multicolumn{1}{c|}{Versicolor} &        & 1.000  & 1.000  & 1.000  \\ \cline{3-3}
\multicolumn{1}{c|}{} & \multicolumn{1}{c|}{}     & \multicolumn{1}{c|}{Virginica}  &        & 1.000  & 1.000  & 1.000  \\ \cline{2-7} 
\multicolumn{1}{c|}{} &
  \multicolumn{1}{c|}{\multirow{3}{*}{QSVM}} &
  \multicolumn{1}{c|}{Setosa} &
  \multirow{3}{*}{0.6333} &
  0.667 &
  0.571 &
  0.640 \\ \cline{3-3}
\multicolumn{1}{c|}{} & \multicolumn{1}{c|}{}     & \multicolumn{1}{c|}{Versicolor} &        & 0.375  & 0.666  & 0.500  \\ \cline{3-3}
\multicolumn{1}{c|}{} & \multicolumn{1}{c|}{}     & \multicolumn{1}{c|}{Virginica}  &        & 0.900  & 0.636  & 0.783  \\ \hline
\end{tabular}
\end{table}

%% file: tex/future.tex
\label{future}

As we pointed out in the earlier chapters,  Quantum Machine Learning is a rapid-growing field. We have tried to work with a small part of this gigantic field. This field is still recent. There are lots of opportunities for the researchers in this field. In our investigations, we have tried to explore two of the most promising algorithms in quantum machine learning sector.

We have studied the HHL algorithm in the perspective of a matrix being diagonal or non-diagonal, having different sparsity and of different shapes. There are other parameters for this algorithm too, which can affect the fidelity and the probability of the solutions. We believe that the researchers will concentrate on these issues and try to figure out valuable insights.

QSVM section also has considerable research to do. We have worked with two standard datasets. But real world datasets can be more diverse and challenging to fit in. To create real impact on the world, we need to work with real life datasets. We are looking forward to continue our work too.

We have so far focused on the number of shots to increase the accuracy. But there are other issues which can be focused as well. We used QSVM as our model. Our machine learning and deep learning models are waiting to be implemented and benchmarked on some quantum devices. Moreover, we have implemented the linear kernel of the SVM algorithm. We hope many researchers will also work with other kernel tricks. 

In a nutshell, there are so many windows yet to be opened. We hope our work can initiate anyone's journey to the world of quantum computing and help others to find something valuable for humankind.

%% file: tex/conclusion.tex
\label{conclusion}
In this paper, We have presented two quantum subroutines of machine learning algorithms for the purpose of solving system of linear equations and for supervised classification of data.
\par The Harrow-Hassidim-Lloyd algorithm is designed to find out the solution for a system of the linear equations on a quantum device. In our contribution, we wanted to investigate how effectively this algorithm performs when the environment is unusual. We tried to study and explain the background of the HHL algorithm in brief before moving forward to further contribution and we have been able to detect some limitations. We have experimented with diagonal and non-diagonal matrices of  of  $2\times2$, $4\times4$ and $8\times8$ dimensions. Further, we have also analyzed the algorithm with arbitrary sparse hermitian matrices of different densities such as $0.5$ and $1$.  We started our experiment by mentioning two hypotheses. We found one of them to be true. We suspected that the HHL algorithm would perform better on diagonal matrix than on non-diagonal matrix. After experimenting we have found that this algorithm demands more processing power to process the non-diagonal matrices than the diagonal ones. To solve the system with non-diagonal matrices, it has produced deeper and wider circuits and has taken more processing time. We also presumed that this algorithm would perform better on lower density random hermitian matrices than higher density ones but this assumption was not accurate in this case. Because this algorithm failed to perform well in both situations. We have discovered that this algorithm produces invalid results when we provide random hermitian matrices. Those outputs are not trustworthy at all as the differences with the real outputs are immense. There are some positive outcomes too. This algorithm performs almost accurately with diagonal and non-diagonal matrices. Another positive thing is it does not require bigger circuit to solve higher density matrices. Our article shows the necessity of tuning this HHL algorithm or finding a new algorithm which can survive these critical cases.  
\par As quantum computing is comparatively a modern research field, many noble contributions are yet to be done. Enthusiasts can explore how the HHL algorithm performs when we provide larger matrices and with different densities. We hope that this contribution may open a new window to explore HHL and such algorithms.
\par Quantum Suppot Vector Machine is also a promising algorithm in the field of Quantum Machine Learning. We have experimented with two datasets named Breast Cancer dataset and Iris Dataset. QSVM with linear kernel and SVM with linear kernel have been run on both of these datasets. We have pre-processed the data, applied Principal Component Analysis on the datasets, normalized them around $0$ and scaled them between $-1$ and $+1$. Though QSVM has not performed as well as SVM performed on both of the datasets, there are so many possibilities to increase the efficiency of this algorithm. 
Our research has been a small contribution to the vast field of quantum machine learning.  We hope that more research work will be conducted on this field and they will produce more valuable knowledge.

%% file: tex/acknowledgement.tex
We would like to thank the Department of Computer Science and Engineering, Shahjalal University of Science and Technology, Sylhet 3114, Bangladesh for supporting this research work. We are also thankful to the authors of previous works for their excellent contribution. We would like to express our gratitude to our respected supervisors Summit Haque and Dr. Omar Shehab for their massive guidance, assistance and encouragement throughout the journey. 

%% file: tex/dedication.tex
We would like to dedicate this research work to our beloved parents and to our younger sisters.